\documentclass[a4paper,11pt]{article}
\pdfoutput=1 % if your are submitting a pdflatex (i.e. if you have
\usepackage{jheparxiv}

\usepackage[T1]{fontenc} % if needed

\usepackage{tensor}
\usepackage{amsmath}
\usepackage{amssymb}
\usepackage{mathrsfs}
\usepackage{mathtools}
\usepackage{bbm}
\usepackage{graphicx}
\usepackage{stmaryrd}
\usepackage{twistor}
\usepackage{subcaption}
\usepackage{xcolor}
\usepackage{empheq}

\newcommand{\qed}{\hfill \ensuremath{\Box}}

\newcommand{\eps}{\epsilon}
\newcommand{\veps}{\varepsilon}

\newcommand{\CO}{\mathscr{O}}

\renewcommand{\sc}{\mathsf{c}}

\title{Scattering on self-dual Taub-NUT}

\author[a]{Tim Adamo,}
\author[b]{Giuseppe Bogna,}
\author[b]{Lionel Mason,}
\author[c]{Atul Sharma}

\affiliation[a]{School of Mathematics and Maxwell Institute for Mathematical Sciences,\\
University of Edinburgh, EH9 3FD, UK\vspace{0.1cm}}
\emailAdd{t.adamo@ed.ac.uk}

\affiliation[b]{The Mathematical Institute, University of Oxford,\\ Woodstock Road, Oxford OX2 6GG, United
Kingdom\vspace{0.1cm}}
\emailAdd{giuseppe.bogna@maths.ox.ac.uk}
\emailAdd{lmason@maths.ox.ac.uk}

\affiliation[c]{Center for the Fundamental Laws of Nature \& Black Hole Initiative,\\Harvard University, Cambridge, MA, 02138, USA \vspace{0.1cm}}
\emailAdd{atulsharma@fas.harvard.edu}

\abstract{We derive exact solutions of massless free field equations and tree-level two-point amplitudes up to spin 2 on self-dual Taub-NUT spacetime, as well as on its single copy, the self-dual dyon. We use Killing spinors to build analogues of momentum eigenstates, finding that, in the spirit of color-kinematics duality, those for the self-dual dyon lift directly to provide states on the self-dual Taub-NUT background if one replaces charge with energy. We discover that they are forced to have faster growth at infinity than in flat space due to the topological non-triviality of these backgrounds. The amplitudes for massless scalars and spinning particles in the $(+\,+)$ and $(+\,-)$ helicity configurations vanish for generic kinematics as a consequence of the integrability of the self-dual sector. The $(-\,-)$ amplitudes are non-vanishing and we compute them exactly in the backgrounds, which are treated non-perturbatively. It is explained how spin is easily introduced via a Newman-Janis imaginary shift along the spin-vector leading directly to the well-known exponential factor in the dot product of the spin with the momenta. We also observe a double copy relation between the gluon amplitude on a self-dual dyon and graviton amplitude on a self-dual Taub-NUT spacetime.}
\begin{document}

\maketitle
\flushbottom

\section{Introduction}

Physics in black hole backgrounds remains a major avenue for explorations in classical and quantum gravity. One concrete setting which captures both of these regimes is that of \emph{scattering} off a black hole background. In the classical context, the computation of probe or wave scattering off a black hole encodes classical observables including scattering angle, waveform and power emitted (cf., \cite{Sanchez:1976fcl,Sanchez:1977si,Handler:1980un,Futterman:1988ni,Glampedakis:2001cx,Bautista:2021wfy,Kol:2021jjc,Bautista:2022wjf}). Indeed, the computation of scattering amplitudes with emission in curved background spacetimes provides the on-shell building blocks of the self-force expansion~\cite{Adamo:2022rmp,Adamo:2022qci,Adamo:2023cfp,Cheung:2023lnj,Kosmopoulos:2023bwc}, which systematizes corrections to geodesic motion due to radiation reaction (cf., \cite{Poisson:2011nh,Barack:2018yvs}). In the quantum context, scattering near black hole event horizons is a playground where emergent features of quantum gravity in strongly curved spacetimes can potentially be studied~\cite{tHooft:1996rdg,Betzios:2016yaq,Goldberger:2019sya,Goldberger:2020geb,Goldberger:2020wbx,Betzios:2020xuj,Kallosh:2021ors,Gaddam:2021zka}. More generally, in proposals for holography in asymptotically flat spacetimes, such as celestial holography (cf., \cite{Raclariu:2021zjz,Pasterski:2021raf}), the S-matrix is the natural boundary observable. Scattering amplitudes on black hole spacetimes thus provide key data for asymptotically flat holography.

However, the analytic computation of scattering on black hole spacetimes is rife with difficulties. One obvious complication is the fact that the S-matrix does not exist as a unitary operator in the background QFT, due to the presence of the event horizon. This problem can be circumvented, at least perturbatively, by providing a purely variational definition of tree-level `scattering amplitudes' as multi-linear pieces of the classical background field action, evaluated on recursively constructed solutions to the background-coupled equations of motion. This is sometimes called the `perturbiner' approach~\cite{Arefeva:1974jv,Abbott:1983zw,Jevicki:1987ax,Selivanov:1999as}, and scattering amplitudes defined in this way contain all of the expected dynamical information~\cite{Adamo:2017nia,Ilderton:2023ifn,Kim:2023qbl}. 

This does nothing to ameliorate the underlying technical complication of determining the ingredients necessary to compute a scattering amplitude in a black hole background. This is apparent already at the level of solving the linearised, background-coupled equations of motion that determine the asymptotic scattering states. Even in the toy model of scattering scalar charges on a Coulomb potential, the solutions to the scalar wave equation are given by confluent hypergeometric functions, while on a Schwarzschild metric, solutions of the scalar wave equation are confluent Heun functions (cf., \cite{Bagrov:1990xp}). Linearised gravitational waves (i.e., gravitons) are determined by solving the Regge-Wheeler-Zerilli equations on Schwarzschild~\cite{Regge:1957td,Zerilli:1970se}, or the Teukolsky equations on Kerr~\cite{Teukolsky:1972my,Teukolsky:1973ha}. Similar separability techniques extend to Kerr-Taub-NUT metrics (cf., \cite{Bini:2002kp,Bini:2003sy,Bini:2003bm}), with the linearised fields again determined by a radial equation of Schr\"odinger type. While these equations are amenable to numerical methods, constructing general solutions analytically is hard.  

One possibility for circumventing these (and other) difficulties is to first compute scattering amplitudes on \emph{self-dual} analogues of black hole spacetimes. The underlying integrability of the self-dual sector could -- in principle, at least -- enable all-multiplicity amplitude computations around the self-dual background, which could in turn be leveraged to perturbatively reconstruct the anti-self-dual part of the physical, non-chiral black hole. Of course, this requires first being able to compute all-multiplicity information on a self-dual background in the first place, which is far from obvious. Furthermore, scattering on self-dual backgrounds is of interest in its own right, particularly in the context of celestial holography, where the only known top-down construction of a holographic dual involves an asymptotically flat self-dual (scalar-flat K\"ahler) bulk geometry~\cite{Costello:2022jpg,Costello:2023hmi}.

In~\cite{Adamo:2020syc,Adamo:2020yzi,Adamo:2021bej,Adamo:2022mev} a framework was developed to compute all-multiplicity graviton scattering on self-dual \emph{radiative} spacetimes, as well as all-multiplicity gluon scattering on self-dual radiative gauge fields. These backgrounds are determined by their free radiative data at null infinity. As they are fully non-linear backgrounds, the resulting scattering amplitudes resum infinitely many contributions that would arise in conventional perturbation theory around a trivial vacuum. While this realizes the goal of obtaining all-multiplicity results around the self-dual sector, black hole spacetimes are stationary with no radiation, so do not fall into the radiative class of metrics.

In this paper, we begin the task of extending the work of~\cite{Adamo:2020syc,Adamo:2020yzi,Adamo:2021bej,Adamo:2022mev} to self-dual black hole spacetimes. The first order of business in this regard is to determine the massless free fields, the analogues of momentum eigenstates, on such backgrounds. These will serve as the external wavefunctions for massless scattering processes on self-dual black hole backgrounds. Furthermore, knowing them immediately allows us to determine the 2-point amplitudes on the background, by simply evaluating the quadratic part of the background field action on two such free fields.

The paradigmatic self-dual black hole is the \emph{self-dual Taub-NUT} (SDTN) solution~\cite{Hawking:1976jb}. This is the unique complexified metric in the Taub-NUT family~\cite{Taub:1950ez,Newman:1963yy} that is self-dual, with the mass $M$ and NUT parameter $N$ related by $N=-\im M$. The SDTN metric is also the self-dual part of the Kerr metric, as can be seen by analytically continuing through an imaginary shift in the direction of the spin vector. This is the basis of the Newman-Janis correspondence~\cite{Newman:1965tw} and is particularly easy to see by expressing the ASD Weyl spinor, $\Psi_{\alpha\beta\gamma\delta}$, in terms of the valence-two Killing spinor, $\chi_{\alpha\beta}$ (cf., \cite{Walker:1970un,Hughston:1972qf,Hughston:1973, Jeffryes:1984a} and \S6 of \cite{Penrose:1986uia}). In particular,
\begin{equation}
\Psi_{\alpha\beta\gamma\delta}=M\,\frac{\chi_{(\alpha\beta}\,\chi_{\gamma\delta)}}{\chi^5}    \, , \qquad \chi^2:=\chi_{\alpha\beta}\chi^{\alpha\beta}\, , \label{weyl-KS}
\end{equation}
where $M$ is the mass and in the rectangular coordinates $\vec{x}$ for the original Kerr metric~\cite{Kerr:1963ud}, $\chi_{\alpha\beta}=(\vec{x}-\im\vec{a})\cdot\vec{\sigma}_{\alpha\beta}$, for $\vec{\sigma}_{\alpha\beta}$ the 3-vector of Pauli matrices and $\vec{a}$ the spin vector aligned along the $z$-axis. The self-dual Weyl spinor is obtained from the complex conjugate $\bar \chi_{\dot\alpha\dot\beta}$. Both Weyl spinors are obtained from the corresponding expressions for the Schwarzschild metric by opposite imaginary shifts along $\pm \vec{a}$.  For the full Schwarzschild and Kerr metrics, this requires a somewhat mysterious nonlinear superposition of shifts in both  $\pm \im\vec{a}$.  However, in the  self-dual sector, the spinning SD Taub-NUT metric is simply the complex metric obtained by translating the original SDTN along $\im\vec{a}$.  We will use this fact later to obtain spinning analogues of our formulae.

\medskip

Our focus in this paper will be on constructing solutions to the massless free field equations of spin 0, 1 and 2 on the SDTN background, and computing the resulting two-point scattering amplitudes. In the spirit of the double copy~\cite{Luna:2015paa}, we also perform the same computations for the \emph{self-dual dyon} (SDD), the self-dual combination of the Coulomb solution and a magnetic monopole. Indeed we see that our momentum eigenstates on the SDD background lift directly to provide those on the SDTN background.

We obtain explicit expressions for linear, massless background-coupled fields on SDD and SDTN backgrounds. In the study of self-dual radiative backgrounds, linear fields were determined using twistor methods~\cite{Adamo:2020yzi,Adamo:2022mev}, but on the SDD and SDTN backgrounds more elementary techniques can be used\footnote{Indeed, in contrast to the radiative case, SDD and SDTN do not have functional degrees of freedom.}. Our main tool for constructing our candidate momentum eigenstates is to introduce novel \emph{charged Killing spinors}.  There is then an elementary lift from charged massless fields on SD dyons to massless fields on SD Taub-NUT, with the energy of the latter related to the charge of the former. These are special  structures on  these backgrounds and result in remarkable, elementary expressions that are \emph{far} simpler than the confluent hypergeometric functions expected in the Coulomb case, or the Heun functions for Schwarzschild or Kerr. This simplicity allows us to compute the 2-point functions of these massless linear fields by direct integration.

Another novelty associated to these computations, which can be viewed in both a positive and negative light, is that both the SDD and SDTN are \emph{topologically non-trivial}. The SDD is the self-dual superposition of the Coulomb and Dirac monopole fields; the latter has a non-trivial Chern class on the spheres of constant radius and the monopole charge is necessarily quantized~\cite{Dirac:1931kp,Wu:1975es,Wu:1976ge,Shnir:2005vvi}. Charged fields therefore have spin weight on the sphere, and consequently momentum eigenstates have non-trivial spherical harmonics, leading to higher radial growth at infinity than expected for un-charged fields. Such changes in asymptotic boundary conditions are an expected feature due to the underlying monopole topology (cf., \cite{Tamm:1931dda,Banderet:1946,Ford:1959,tHooft:1974kcl,Schwinger:1976fr,Boulware:1976tv}).

A similar phenomenon arises on SDTN, where the topology enforces a periodic time variable leading to a circle bundle with non-trivial topology on spheres of constant radius~\cite{Newman:1963yy,Misner:1963fr}. This leads to quantization of the energy $\omega$ of the particles $4M\omega\in\Z$, and fields with non-vanishing energy have non-trivial spin weight on the sphere. Again, this means that momentum eigenstates on the SDTN spacetime have enhanced growth at infinity.

One could worry that this enhanced growth at infinity renders the scattering amplitudes of these background-coupled fields ill-defined. Indeed, the usual way of computing 2-point amplitudes in point-like potentials is to extract the leading $r^{-1}$ behaviour of the free fields as $r\to\infty$ (cf., \cite{Taylor:1972pty}), or evaluate the quadratic action as a large-$r$ boundary term~\cite{Adamo:2021rfq,Adamo:2023cfp}, and this perspective is complicated somewhat by the non-trivial topology reflected in our solutions. We find it more natural to evaluate the 2-point amplitude by direct integration of the quadratic action, evaluated on a superposition of two of the free field solutions. 

Remarkably, the enhanced polynomial growth in $r$ is compensated by the plane wave phase in such a way that we obtain simple, finite results. For instance, negative helicity graviton scattering on the spinning SD Taub-NUT background is given by 
\be
%\boxed{\hspace{1em}
\mathcal{A}(1,2) = 4\pi^2\,M\,(4M\omega_2)!\,\e^{-(\vec{k}_1+\vec{k_2})\cdot \vec{a}}\,\frac{(2\omega_1)^{4M\omega_2-2}\,\la 1\,2\ra^{4+4M\omega_2}}{|\vec{k}_1+\vec{k}_2|^{8M\omega_2+2} }\,\delta_{-\omega_1,\omega_2}\,,
%\hspace{1em}}
\label{eq:integrated-2-point-graviton-spin}
\ee
where the particles have null momenta $(\omega_i,\vec{k}_i)$ with energies $\omega_i=\pm |\vec{k}_i|$, quantized as above and $\langle 1\,2 \rangle$ is the usual spinor-helicity contraction of polarization spinors for the two particles. The  factor of $\e^{-(\vec{k}_1+\vec{k_2})\cdot \vec{a}}$  follows from \eqref{eq:integrated-2-point-graviton} by translating the contour used in the non-spinning case (see \eqref{eq:integral-2-point-graviton} below) by $\im\,\vec{a}$ as described above.   Here, the anomalous little group weight arises from the topological non-triviality of the background, and can be removed by a suitable normalization of the states, as described below. A similar story holds for gluon scattering on a SD dyon background (embedded in a Cartan subalgebra of the non-abelian gauge group). 

\medskip

\paragraph{Summary of the paper:} In section \ref{sec:backgrounds}, we provide brief reviews of the SD dyon and SD Taub-NUT backgrounds. Section \ref{sec:states} starts by introducing the main tool, charged solutions to the Killing spinor equation coupled to the SD dyon. These allow us to make general solutions on flat space to charged solutions in the self-dual dyon background. This in turn enables the introduction of general momentum eigenstates and solutions to the Hamilton-Jacobi equations on the background. We then show how these solutions can be lifted directly to the SD Taub-NUT background.  Various other solutions of different helicity and in potential form can then be found.
Section \ref{sec:amps} is concerned with deriving 2-point tree-level around the backgrounds under consideration. Section \ref{sec:conclusions} contains a discussion of the interpretation, the extension to a spinning SDTN background, the connection to flat space amplitudes, as well as discussion of possible directions for future research. Appendix \ref{app:int} contains details of the integrals involved in the evaluation of the 2-point amplitudes.

%%%%%%%%%%%%%%%%%%%%%%%
%%%%%%%%%%%%%%%%%%%%%%%

\section{The classical backgrounds}
\label{sec:backgrounds}

We begin by reviewing our classical backgrounds of interest, starting with the self-dual dyon (SDD) background in Euclidean signature. This starts life as a background in electrodynamics, but can also be embedded as a Cartan-valued background in non-abelian gauge theories. We then describe the metric of self-dual Taub-NUT (SDTN) space in Lorentzian, Euclidean and split signatures. In the following sections, we will focus on Euclidean signature.

%%%%%%%%%%%%%%%%%%%%%%%

\subsection{Self-dual dyon}
\label{sec:sddyon}

The self-dual dyon (SDD) arises, in the first instance, as a classical solution of abelian gauge theory. A general dyon profile solving Maxwell's equations on Euclidean $\R^4$ can be constructed through a superposition of electric and magnetic charges. The self-dual (SD) profile has equal electric and magnetic charges and is captured by the gauge field
\be
A = \frac{\d t}{r} + a\, , \qquad a=  (1-\cos\theta)\,\d\phi\,.\label{eq:abelian-sd-dyon}
\ee
Here $(r,\theta,\phi)$ are standard spherical polar coordinates on space $\R^3-0=\R_+\times S^2$, and $t$ is the Wick-rotated time coordinate from the Lorentzian slice. 

The first term in \eqref{eq:abelian-sd-dyon} is the Coulomb part of the potential sourced by the electric charge of the dyon, whereas the second term $a$ is the Dirac monopole sourced by the magnetic charge. This has a wire singularity at $\theta=\pi$ which can be moved to $\theta=0$ by the gauge transformation $\e^{2\im\phi}$. The associated field strength 
\be
F = \d A = \frac{\d t\wedge\d r}{r^2} + \sin\theta\,\d\theta\wedge\d\phi
\ee
is smooth on the sphere and is easily seen to be self-dual with respect to the Hodge star operator associated to the flat Euclidean metric on $\R^4$. As such, it satisfies Maxwell's equations and the electric and magnetic charge are equal.

In order to construct non-trivial amplitudes on this background, we can embed the SDD into the Cartan subalgebra of the gauge algebra of a generic non-abelian Yang-Mills theory. Let $\sc$ be an element of the Cartan subalgebra and simply set
\be\label{sdc}
A = \sc\left(\frac{\d t}{r} + a\right)\,.
\ee
For example, in the simplest case of $\SU(2)$ the element $\sc$ can be taken to be the Pauli matrix $\sigma_3$. Since the gauge field lies in an abelian subalgebra of the gauge algebra, the corresponding field strength reduces to
\be
F=\d A+\frac12\,[A,A]=\sc\left(\frac{\d t\wedge\d r}{r^2} +\sin\theta\,\d\theta\wedge\d\phi\right)\,.
\ee
This is essentially the same as the abelian field strength, and continues to satisfy the (non-abelian) self-duality equations.

%%%%%%%%%%%%%%%%%%%%%%%%%%%%%%%%%%%%%%%%%%%%%%

\subsection{Self-dual Taub-NUT}
\label{sec:sdtn}

\paragraph{Lorentzian signature:}
The Lorentzian-real Taub-NUT metric takes the form
\begin{equation}
    \d s^2= f(r)\left(\d t-2N\,a\right)^2- \frac{\d r^2}{f(r)} -\left(r^2+N^2\right) \d\Omega^2\, ,
\end{equation}
where $a$ is the monopole field from \eqref{eq:abelian-sd-dyon} and
\begin{equation}
 f(r)= \frac{r^2-2Mr-N^2}{r^2+N^2}=1- \frac{M+\im N}{r+\im N} -\frac{M-\im N}{r-\im N}\, . \end{equation}
Here $M$ is ADM mass and $N$ is the NUT charge; $\d\Omega^2$ is the round sphere metric on $S^2$, whereas $\d a$ is the volume form on $S^2$. 

In a choice of complex stereographic coordinate $\zeta=\e^{\im\phi}\tan(\theta/2)$ one has 
\begin{equation}
\d\Omega^2= m\, \bar m\, , \qquad m\coloneqq\frac{2\,\d\zeta}{1+|\zeta|^2}\,.
\end{equation}
Gauge transformations of $a\rightarrow a +\d g$, for some function $g$, are accompanied by coordinate transformations $t\rightarrow t+2Ng$. Globally, this requires the identification  $t\sim t +8\pi \im N$.
Setting $N=-\im M$ produces the complex metric
\begin{equation}
    \d s^2= f(r)\,(\d t+2\im\,Ma)^2- \frac{\d r^2}{f(r)} -\left(r^2-M^2\right) \d\Omega^2\, ,\qquad f(r)=\frac{r-M}{r+M}\, , \label{Lorentz-SDTN}
\end{equation}
which is vacuum and self-dual (in the holomorphic, complexified category).

\paragraph{Euclidean signature:} The metric \eqref{Lorentz-SDTN} acquires a Euclidean real slice with the replacement of $t$ by $\im t$, and $r-M$ by $r$, taking $M$ to be real~\cite{Hawking:1976jb}.  Setting 
\be
\vec{x} = (r\sin\theta\cos\phi,\,r\sin\theta\sin\phi,\,r\cos\theta)\,,\qquad (\theta,\phi)\in S^2\,,
\ee
the SDTN metric takes the form of a Gibbons-Hawking gravitational instanton~\cite{Gibbons:1978tef,Gibbons:1979xm}:
\be\label{sdtn}
\begin{aligned}\d s^2 &=\left(1+\frac{2M}{r}\right)^{-1}\left(\d t-2Ma\right)^2+\left(1+\frac{2M}{r}\right)\d \vec x^2\\&\equiv V^{-1}(\d t-2Ma)^2 + V\,\d\vec{x}^2\,,\end{aligned}
\ee
where $\d\vec{x}\,^2 = (\d x^1)^2+(\d x^2)^2+(\d x^3)^2$, and $V$, $a$ are given by
\be\label{Vomega}
V \coloneqq 1+\frac{2M}{r}\,,\qquad a \coloneqq (1-\cos\theta)\d\phi\,.
\ee
More generally, this ansatz gives a self-dual Ricci-flat metric when the scalar and vector potentials $(V, a)$ solve the Bogomolny equation on $\R^3-0$,
\be
\p_iV + 2M\eps_i{}^{jk}\p_ja_k = 0\,,
\ee
where $\p_i\equiv\p/\p x^i$, $i=1,2,3$. With $(V,a)$ given by \eqref{Vomega}, they describe the Euclidean SDTN metric.

Taking $r\geq 0$ and identifying $t\sim t+8\pi M$, this gives a complete, non-trivial, self-dual Ricci-flat metric on $\R^4$. To see this, first combine the spherical coordinates $(\theta,\phi)\in S^2$ with the periodic coordinate $t\in S^1$ to find the total space of the Hopf fibration  $S^3\rightarrow S^2$ with fiber $S^1$. Together with $r\in \R_+$ these make up coordinates on $\R^4-0=\R_+\times S^3$.  One then completes the spacetime into $\R^4$ by adding back the origin to $\R^4-0$. The $S^3$s at constant $r$ approach spheres of radius $\sqrt {2Mr}$ as $r\rightarrow 0$, and shrink to a point at the origin~\cite{lebrun1991complete}. 

The locus $r=0$ is a Euclidean analogue of the black hole horizon for SDTN. However, in Euclidean signature this is not an actual horizon, but only a single point. The metric has an apparent singularity at $r=0$, but can be extended across it by a standard change of coordinates. Computing curvature invariants like the Kretschmann scalar, one finds a true curvature singularity at $r=-2M$, but for $M>0$, this point is outside of Euclidean $\R^4$. Equivalently, it can be thought of as being hidden behind the `horizon'. This intuition becomes precise in split signature, as we briefly mention below.

The metric \eqref{sdtn} is easily verified to be Ricci-flat. For the orientation in which the volume form is $\sqrt{|g|}\,\d t\wedge\d x^1\wedge\d x^2\wedge\d x^3$, \eqref{sdtn} is also self-dual. That is, the Weyl tensor $C_{\mu\nu\rho\sigma}$, viewed as a $2$-form on either its first two or last two indices, is self-dual: $C=*C$, where $*$ is the Hodge star of \eqref{sdtn} in the chosen orientation. This means that the metric is hyperk\"ahler: there exist a 2-sphere's worth of complex structures on $\R^4$ that are compatible as well as covariantly constant with respect to this metric.

\paragraph{Split signature:} The global structure of the SDTN spacetime in split (i.e., $(+-\,-\,+)$) signature was studied in~\cite{Crawley:2021auj,Crawley:2023brz}. In this signature, the authors of~\cite{Crawley:2021auj} interpret it as a genuine black hole geometry.

The Wick rotation that takes \eqref{sdtn} from Euclidean to split signature while preserving the sign of the volume form $\sqrt{|g|}\,\d t\wedge\d x^1\wedge\d x^2\wedge\d x^3$ is given by $(x^1,x^2)\mapsto(\im x^1,-\im x^2)$, or equivalently $\theta\mapsto\im\theta,\phi\mapsto-\phi$. The resulting metric reads\footnote{The metric employed in \cite{Crawley:2021auj} is found by a further change of coordinates $t\mapsto t-2M$, $r\mapsto r-M$.}
\be\label{sdtnsplit}
\d s^2 = \left(1+\frac{2M}{r}\right)^{-1}\big[\d t+2M\,(1-\cosh\theta)\,\d\phi\big]^2 + \left(1+\frac{2M}{r}\right)\d\vec{x}^2\,,
\ee
where now $\d\vec{x}^2 = (\d x^3)^2-(\d x^1)^2-(\d x^2)^2$. In this signature, the coordinates range over $r,\theta\geq0$, $\phi\sim\phi+2\pi$, $t\sim t+8\pi M$.

Once again, the spacetime has a coordinate singularity at $r=0$. This happens to be a genuine horizon instead of a single point as in Euclidean signature, and the metric can be extended beyond this horizon into the interior of this black hole by continuing $r$ to the negative reals. The maximal extension encounters a curvature singularity at $r=-2M$, where the spacetime ends. This justifies the use of the terminology \emph{self-dual black hole}.

Another important distinction between the geometries is that in Euclidean signature, surfaces of constant $t,r$ are spatial 2-spheres parametrized by $\theta,\phi$. In split signature, a surface of constant $t$ and constant $r^2=(x^3)^2-(x^1)^2-(x^2)^2$ is a two-dimensional hyperbolic space $\HH_2$, also known as the Poincar\'e disk. So, for $r>0$, the metric can be thought of as a metric on a circle bundle over $\R_+\times \HH_2$.

%%%%%%%%%%%%%%%%%%%%%%%
%%%%%%%%%%%%%%%%%%%%%%%

\section{Massless free fields from Killing spinors}
\label{sec:states}

The first step in understanding quantum field theory on curved spacetimes is to construct the Hilbert space of states. To study perturbation theory around the free limit, this is taken to be the space of solutions of the classical free field equations. In Euclidean signature, these are the Laplace equation and its higher spin counterparts. In this section, we show that it is possible to find exact expressions for classical solutions of massless free field equations on both the SD dyon and SD Taub-NUT backgrounds. This is in stark contrast to the corresponding situation in real Coulomb potentials, physical black holes like Schwarzschild or Kerr, or generic members of the Kerr-Taub-NUT family, where such solutions can only be found as spherical harmonic expansions with radial wavefunctions given by complicated special functions (cf., \cite{Bagrov:1990xp}). 

Our solutions will instead be plane waves dressed with simple power laws, reducing to familiar momentum eigenstates in the flat limit $M\to0$ or $\sc\to0$. For this reason, we refer to these solutions as \emph{quasi}-momentum eigenstates, as there is no actual translation invariance in the SDD or SDTN backgrounds. We first construct scalar and negative helicity states, and then describe a spin-raising procedure to construct positive helicity states. We mainly focus on spin 1 states around the SD dyon and spins 1 and 2 around SD Taub-NUT. Our main technique will be to introduce charge-raising and lowering Weyl spinors that solve certain Killing spinor equations.\footnote{There is considerable literature on Dirac Killing spinors on Taub-NUT and related geometries, see for example~\cite{Martelli:2011fu,Martelli:2011fw,Martelli:2012sz,Martelli:2013aqa,Toldo:2017qsh}. It would be interesting to study their relation, if any, to our Weyl Killing spinors.} These allow us to turn standard flat space momentum eigenstates into states on SDD and SDTN backgrounds.

\paragraph{Spinor-helicity conventions:} In what follows, it will be useful to work with spinor-helicity variables $\kappa^\alpha$ and $\tilde\kappa^{\dot\alpha}$ parametrizing null momenta $k^\mu$ via the decomposition
\be
k^{\al\dal} = \frac{1}{\sqrt{2}}\begin{pmatrix}\omega+\im k^3&& k^2+\im k^1\\
- k^2+\im k^1&&\omega-\im k^3\end{pmatrix} = \kappa^\al\tilde\kappa^{\dal}\,.
\ee
Here, $\alpha=0,1$ and $\dot\alpha=\dot 0,\dot 1$ are left- and right-handed 2-component spinor indices, respectively. One can express $\kappa^\alpha$ and $\tilde\kappa^{\dot\alpha}$ in terms of stereographic coordinates $(z,\tilde z)\in\CP^1\times\CP^1$ on the complexified celestial sphere,
\be
z := \frac{k^{1\dot0}}{k^{0\dot0}} = \frac{-k^2+\im k^1}{\omega+\im k^3}\,,\qquad \tilde z := \frac{k^{0\dot1}}{k^{0\dot0}} = \frac{k^2+\im k^1}{\omega+\im k^3}\,.
\ee
Up to little group scalings $(\kappa,\tilde\kappa)\sim(s\kappa,s^{-1}\tilde\kappa)$, $s\in\C^*$, we may solve for the spinor-helicity variables to find
\be
\kappa^\al = (1,z)\,,\qquad\tilde\kappa^{\dal} = \frac{\sqrt{2}\,\omega}{1+z\tilde z}\,(1,\tilde z)\,.\label{eq:lambdas}
\ee
Fixing $\kappa^0=1$ as done here happens to be a convenient choice in self-dual backgrounds. The Levi-Civita invariants $\veps^{\al\beta},\veps^{\dal\dot\beta}$ of $\SL(2,\C)$ are used to raise and lower spinor indices and to construct spinor brackets in the standard way: $\langle \lambda\,\kappa\rangle\coloneqq\varepsilon^{\alpha\beta}\lambda_\beta\kappa_\alpha=\lambda^\alpha\kappa_\alpha$ and $[\tilde\lambda\,\tilde\kappa]\coloneqq\varepsilon^{\dot\alpha\dot\beta}\tilde\lambda_{\dot\beta}\tilde\kappa_{\dot\alpha}=\tilde\lambda^{\dot\alpha}\tilde\kappa_{\dot\alpha}$ (using the conventions of~\cite{Adamo:2022mev}).

Similarly, we can introduce the spinor equivalent $x^{\al\dal}$ of the spacetime coordinates $x^\mu\in\R^4$. In Euclidean signature, this is given by
\be
x^{\al\dal} = \frac{1}{\sqrt{2}}\begin{pmatrix}
    x^0+\im x^3&&x^2+\im x^1\\-x^2+\im x^1&&x^0-\im x^3
\end{pmatrix}\,.
\ee
In our applications, we will take $x^0=t$ to be the periodic coordinate, and $x^i$ to be the spatial coordinates. Similarly, spinor equivalents of tensors on a 4-manifold with Riemannian metric $g$ can be locally defined by working in a tetrad basis $\theta^a$. The spinor equivalent of $\theta^a$ is
\be
\theta^{\al\dal} = \frac{1}{\sqrt{2}}\begin{pmatrix}
    \theta^0+\im \theta^3&&\theta^2+\im \theta^1\\-\theta^2+\im \theta^1&&\theta^0-\im \theta^3
\end{pmatrix}\,.
\ee
The spinor components $A_{\al\dal}$ of any 1-form $A$ are the components in its expansion $A=A_{\al\dal}\theta^{\al\dal}$ in this co-frame, and so on for other tensors; see~\cite{Penrose:1984uia,Penrose:1986uia} for an extensive review of this notation. In flat space, one simply sets $\theta^{\al\dal}=\d x^{\al\dal}$.

%%%%%%%%%%%%%%%%%%%%%%%%%%%%%%%%

\subsection{Killing spinors and raising/lowering charge and helicity}

A valence-$p$ anti-self-dual (ASD) Killing spinor on a general curved spacetime is a solution $\chi^{\alpha_1\ldots \alpha_p}$ to the equation
\begin{equation}
    \nabla^{(\alpha}{}_{\dot\alpha}\chi^{\beta_1\ldots \beta_p)}=0\, .\label{KS}
\end{equation}
We will be interested in ASD Killing spinors of valence $p=1,2$ on SD backgrounds and their helicity raising and lowering properties.

As shown in \S6.4 of~\cite{Penrose:1986uia}, a Killing spinor can be used to raise the helicity of a solution to the helicity $-(n+p)/2$ massless free field equations
\begin{equation}
    \nabla^{\al\dal}\psi_{\alpha\alpha_2\ldots\alpha_{n+p}}=0
\end{equation}
to generate a solution
\begin{equation}
\vphi_{\alpha_1\ldots\alpha_{n}}=\psi_{\alpha_1\ldots\alpha_{n+p}}\,\chi^{\alpha_{n+1}\ldots\alpha_{n+p}}
\end{equation}
of the helicity $-n/2$ field equations. That this satisfies the massless field equations follows directly from the Leibnitz rule and \eqref{KS}. Observe that constant spinors give trivial examples of this, but our focus will be on non-trivial valence 1 and 2 Killing spinors, with those of valence one also raising and lowering charge.

\medskip

The valence-2 ASD Killing spinor appropriate to the SD dyon is given on flat space by choosing a unit timelike vector $T_{\alpha\dot\alpha} = \frac{1}{\sqrt{2}}\diag\,(1,1)$ and writing
\begin{equation}
    \chi^{\alpha\beta}=\sqrt 2 \,T^{(\alpha}{}_{\dot\alpha}x^{\beta)\dot\alpha}\, .
\end{equation}
%With this, $\chi=r$.
Our charge raising and lowering operators will be spinors $\chi^\alpha_\pm$ subject to the relations
\begin{equation}
    \chi^{\alpha\beta}=\chi_+^{(\alpha}\chi_-^{\beta)}\, , \qquad \chi_+^\alpha=\sqrt{2}\,\bar\chi_-^{\dot\alpha}\,T^\alpha{}_{\dot\alpha}\,,
\end{equation}
(or, equivalently, to be complex conjugates in Euclidean signature) which only defines them up to a phase. This phase cannot be chosen to be global on the sphere as the line subbundles they span inside the 4d spin bundle are non-trivial on the sphere.  These line bundles are the spin bundle of the 2-sphere and its dual: $\CO(\mp1)\to S^2$, having  Chern class $\mp 1$, respectively.  These are the monopole line bundles of minimal degree. However, as we will see below, when understood as sections of line bundles of the appropriate Chern class $\pm1$, the spinors $\chi^{\alpha}_{\pm}$ are indeed global on the sphere. 

We can make explicit choices by first defining stereographic coordinates on  the 2-spheres at constant $t$ and $r$
\be
\zeta\coloneqq\frac{x^1+\im x^2}{r+x^3}=\e^{\im\phi}\tan\frac{\theta}{2}\,,\qquad\qquad\bar\zeta=\frac{x^1-\im x^2}{r+x^3}\coloneqq e^{-\im \phi}\tan\frac{\theta}{2}\, .
\ee
With these coordinates, we can take
\begin{equation}\label{chiparam}
   \chi^\alpha_+=\sqrt{\frac{r}{1+|\zeta|^2}}\;(\bar\zeta,-1)
   \, , \qquad \chi^\al_- =\sqrt{\frac{r}{1+|\zeta|^2}}\;(1,\zeta)
\end{equation}
Our main result in this subsection is then:
\begin{propn}
For the self-dual dyon in the gauge
\be
A = \frac{\d t}{r} + a\,,\qquad a = \frac{\im\,(\zeta\d\bar\zeta-\bar\zeta\d\zeta)}{1+|\zeta|^2} = (1-\cos\theta)\d\phi\,,
\ee
satisfying $F_A=*F_A$ with respect to the flat metric $\d s^2 = \d t^2+\d\vec{x}^2$, the spinors $\chi_\pm^\alpha$ satisfy the charged Killing spinor equations
\be\label{Prop3eq}
\p^{(\al}{}_{\dot\alpha}\chi_\pm^{\beta)} \pm\frac{\im}{2}\,A^{(\al}{}_{\dot\alpha}\chi_\pm^{\beta)} = 0\,,
\ee
with the factors of $\pm\im/2$ reflecting the minimal charges associated to the spin bundles of the sphere. Thus, $\chi^{\alpha}_\pm$ are charge $\pm1/2$ Killing spinors. 
\end{propn}

\noindent{}This is proved by direct calculation.  It also follows directly from the twistor treatment of the SD dyon, but we will give that argument elsewhere. 

The SDD potential $A$ and spinors $\chi_\pm^\alpha$ as defined have wire singularities along $\theta=\pi$: the coordinate $\zeta'= 1/\zeta$ is regular at $\theta=\pi$ where $\zeta'=0$, and we find $\chi^\alpha_\pm$ singular there
\begin{equation}
  \chi^\alpha_+=\sqrt{\frac{r}{1+|\zeta'|^2}}\;\left(\frac{|\zeta'|}{\bar\zeta'},-|\zeta'|\right)\, ,\quad\chi^\al_- =\sqrt{\frac{r}{1+|\zeta'|^2}}\;\left(|\zeta'|,\frac{|\zeta'|}{\zeta'}\right)\, .
\end{equation}
However, the respective gauge transformation $\e^{\pm i\phi}=(\zeta/|\zeta|)^{\pm1}=(|\zeta'|/\zeta')^{\pm1}$ removes these singularities % at $\theta=\pi$: 
\begin{equation}
  \chi^\alpha_+=\sqrt{\frac{r}{1+|\zeta'|^2}}\;\left(1,-\bar\zeta'\right)\, ,\quad\chi^\al_- =\sqrt{\frac{r}{1+|\zeta'|^2}}\;\left(\zeta',1\right)\, ,
\end{equation}
at the price of introducing a singularity at $\theta=0$.
Similarly, 
\begin{equation}
\im\,a=\im\,(1-\cos\theta)\,\d\phi\rightarrow -\im\,(1+\cos\theta)\,\d\phi\,,
\end{equation}
under this transformation, which is now regular at $\theta=\pi$.  Thus, as connections and sections of line bundles, everything is globally defined.

The main consequence that we will use is the following:
\begin{corol}\label{corr:3.1}
Let the totally symmetric ASD spinor $\psi_{\alpha_1\cdots \alpha_{n+2e}}$ with $2e\in\Z$ satisfy the flat space wave equation, $\p^{\alpha_1\dot\alpha}\psi_{\alpha_1\cdots \alpha_{n+2e}}=0$. Then 
\begin{equation}
    \vphi_{\alpha_1\cdots \alpha_n} :=\psi_{\alpha_1\cdots \alpha_{n+2e}}\,\chi_+^{\alpha_{n+1}}\cdots \chi_+^{\alpha_{n+2e}}\, ,
\end{equation}
satisfies the helicity $-n/2$ zero-rest-mass equations of charge $e$:
\begin{equation}
    \left(\p^{\alpha_1\dot\alpha}+\im e\,A^{\alpha_1\dot\alpha}\right)\vphi_{\alpha_1\cdots \alpha_n}=0\,.
\end{equation}
When $n=0$ we similarly have $    \left(\p^{\alpha\dot\alpha}+\im e\,A^{\alpha\dot\alpha}\right)    \left(\p_{\alpha\dot\alpha}+\im e\,A_{\alpha\dot\alpha}\right)\vphi=0$.
\end{corol}

\proof This follows for $n>0$ just as in the flat case, because, following an application of the Leibnitz rule, each derivative of $\chi_+^\alpha$ is automatically symmetrized over its undotted indices and comes with one copy of $\frac{\im}{2}\,A$ by \eqref{Prop3eq}. So that $2e$ contractions with $\chi_+^\al$ generate a solution of charge $\frac{1}{2}\cdot 2e=e$.

The case $n=0$ requires a little more work. Write $\D_{\alpha\dot\alpha}=\p_{\alpha\dot\alpha}+\im e\,A_{\alpha\dot\alpha}$ when acting on a field of charge $e$. Taking the case $n=1$, we find
\begin{equation}
    0=\D_{\beta\dot\alpha}\D_{\alpha}{}^{\dot\alpha}\vphi^\alpha = \D^2 \vphi_\beta\,, 
    %+ F_\beta^\alpha \phi_\alpha
\end{equation}
where the first equality follows from the massless equations on $\vphi_\alpha $ and the second from the fact that the term symmetric in $\alpha\beta$ is proportional to the ASD curvature of $A$, which is zero since $A$ is SD. Contracting with a further constant spinor then yields a charged scalar solution to the background-coupled scalar wave equation, and our original $\psi_{\alpha_1\cdots \alpha_{2e}}$ can be taken to arise from a $\psi_{\alpha_1\cdots \alpha_{2e+1}}$ in this way. \qed

\medskip 

Of course, we also have the analogous result for $\chi_-^\alpha$ lowering charge. Note that although the initial valence-2 Killing spinor $\chi^{\alpha\beta}$ can be used to lower helicity, it does not alter the charge, so contracting with $\chi^{\al\beta}$ does not generate any more charged solutions.  

%%%%%%%%%%%%%%%%%%%%%%%%%%%%

\subsection{Scalar and negative helicity states on SD dyon}\label{sec:dyons-KS}

Charged quasi-momentum eigenstates of non-positive helicity on the SDD background can now be generated by beginning with uncharged negative helicity momentum eigenstates on flat space:
\begin{equation}
\psi_{\alpha\cdots\beta}=\kappa_{\alpha}\cdots\kappa_{\beta}\,\e^{\im k\cdot x}\,,
\end{equation}
where the null momentum has spinor decomposition $k_{\alpha\dot\alpha}=\kappa_\alpha \tilde \kappa_{\dot\alpha}$.  Using Corollary~\ref{corr:3.1}, we obtain solutions
\begin{equation}\label{SDDgensol1}
    \vphi_{\alpha_1\cdots\alpha_n} =\kappa_{\alpha_1}\cdots \kappa_{\alpha_n}\,\langle \kappa\, \chi_+\rangle^p\,\langle\kappa\,\chi_-\rangle^q\, \e^{\im k\cdot x}\,,
\end{equation}
of charge $\frac{1}{2}(p-q)$. It is straightforward to check that the charged zero-rest-mass equations still hold for \eqref{SDDgensol1} even when one of $p$ or $q$ is negative, although in this case there will be radial wire singularities where $\la\kappa\,\chi_{\pm}\ra$ appearing in the denominator vanishes.

With the parametrization \eqref{chiparam}, the general quasi-momentum eigenstate of helicity $-n/2$ is given explicitly by
\begin{equation}
    \vphi^{(p,q)}_{\alpha_1\ldots\alpha_n} =\kappa_{\alpha_1}\ldots \kappa_{\alpha_n} \left(\frac{r}{1+|\zeta|^2}\right)^{\frac{p+q}{2}}(\bar\zeta z+1)^p\,(\zeta-z)^q\, \e^{\im k\cdot x}\,,\qquad z\coloneqq\frac{\kappa^1}{\kappa^0}\,,\label{eq:general_wavefunctions}
\end{equation}
where  constant factors of $\kappa^0$ have been removed to give correct little-group weight. A particular feature here is that for smooth solutions on the sphere and non-trivial charge, we see that the $\langle\kappa\chi_+\rangle$ factors generate spin-weighted spherical harmonics as described for example in \S4.15 of \cite{Penrose:1984uia}.  These are, in turn, accompanied by powers of $r$ that, for global solutions on the sphere, are higher than those expected for uncharged fields on a flat background.

In particular, there are distinguished solutions that are nowhere singular on the celestial sphere and grow with the lowest possible power of $r$ as $r\to\infty$. For $n\geq0$, these are given by
\be
\boxed{\hspace{1em}\vphi^+_{\alpha_1\ldots\alpha_n} \coloneqq\vphi_{\alpha_1\ldots\alpha_n}^{(2e,0)}=\kappa_{\alpha_1}\ldots\kappa_{\alpha_n} \bigg(\frac{r}{1+|\zeta|^2}\bigg)^{e}(\bar \zeta z+1)^{2e}\,\e^{\im k\cdot x}\,,\hspace{1em}}\label{eq:phi+}
\ee
for positive quantized charge $2e\in\Z_{\geq0}$, and
\be
\hspace{1cm}\boxed{\hspace{1em}\vphi^-_{\alpha_1\ldots\alpha_n}\coloneqq\vphi_{\alpha_1\ldots\alpha_n}^{(0,-2e)}= \kappa_{\alpha_1}\ldots\kappa_{\alpha_n}\bigg(\frac{r}{1+|\zeta|^2}\bigg)^{-e}(\zeta-z)^{-2e}\,\e^{\im k\cdot x}\,,\hspace{1em}}\label{eq:phi-}
\ee
for negative quantized charge $2e\in\Z_{\leq0}$. The $n=0$ case corresponds to scalar solutions. We refer to these solutions as \emph{minimal fields}, or \emph{minimal states}. 

The other, non-singular solutions can be obtained as derivatives of the minimal states with respect to the external data like momentum.  To see this, write
\be
k\cdot x=\omega\, t+\im \omega\, r\,\frac{(\zeta -z)(\bar\zeta -\tilde z)-(1+\zeta\tilde z)(1+\bar\zeta z)}{(1+|\zeta|^2)\,(1+z\tilde z)}\,,
\ee
which implies that for $p,q\geq0$ we either have
\be
\vphi^{(p,q)}_{\alpha_1\ldots\alpha_n}=\left(\frac{1+z\tilde z}{2\omega}\,\p_{\tilde z}\right)^{q}\vphi^+_{\alpha_1\ldots\alpha_n}\qquad \text{or}\qquad\vphi^{(p,q)}_{\alpha_1\ldots\alpha_n}=\left(\frac{1+z\tilde z}{2\omega}\,\p_{\tilde z}\right)^{p}\vphi^-_{\alpha_1\ldots\alpha_n}\,,
\ee
depending on the sign of the charge $e$. The corresponding amplitudes for these more general states can thus be computed as suitable derivatives of the amplitudes for the minimal states, on which we focus from now on.

Observe that for the minimal scalar states $\vphi^\pm$, the gradients $D_{\alpha\dot\alpha}\vphi^\pm$ define background-dressed null 4-momenta $K^\pm_{\alpha\dot\alpha}(x)=\kappa_\alpha \tilde K^\pm_{\dot\alpha}(x)$ via $\nabla_{\alpha\dot\alpha}\vphi^\pm=\im K^\pm_{\alpha\dot\alpha}\vphi^\pm$, where
\begin{align}
    \tilde K^\pm_{\dot\alpha}(x)&=\tilde\kappa_{\dot\alpha}-\frac{2\im e}{r}\,\frac{\chi_\pm^\alpha\,T_{\alpha\dot\alpha}}{\langle \kappa\, \chi_\pm\rangle}\,.
\label{eq:dressed_momenta_dyon}
\end{align}
Thus, $\vphi^\pm$ satisfy the charged Hamilton-Jacobi equation on the SDD background.

%%%%%%%%%%%%%%%%%%%%%%%

\subsection{Scalar and negative helicity states on SD Taub-NUT}\label{SDTN-KS}

In this subsection we lift our massless fields that are charged with respect to the SD dyon background to provide massless fields background coupled to SD Taub-NUT.  The main idea in fact works for general Gibbons-Hawking metrics.

Here we focus on the ASD zero-rest-mass equations
\begin{equation}
    \nabla^{\alpha\dot\alpha}\psi_{\alpha\alpha_2\ldots \alpha_n}=0\,,
\end{equation}
on the curved SDTN metric. Self-dual Ricci-flatness of the background gives the simplification that the spin connection on the undotted spinor bundle is flat, and can be taken to be identically zero in appropriate frames.  This is the case for the standard co-frame $\theta^a$ of SDTN given by
\be
\theta^{0} = \frac{\d t-2Ma}{\sqrt{V}}\,,\qquad \theta^i = \sqrt{V}\,\d x^i\, ,\label{SDTN-ZRM}
\ee
for which  it can be checked that the ASD 2-forms 
\begin{equation}
\Sigma^i=\theta^0\wedge \theta^i -\varepsilon^i_{jk} \theta^j\wedge \theta^k\,,  
\end{equation}
are closed, leading to the absence of ASD spin coefficients in this frame \cite{Plebanski:1977zz,Capovilla:1991qb}.

Let $e_a=(e_0,e_i)$ be the corresponding dual vector fields.  After multiplying through by $\sqrt{V}$, these are 
\be
\begin{split}
    \sqrt{V}(e_{0},e_i) = (V\p_t\,,\p_i + 2Ma_i \p_t)\qquad
    \implies\qquad\sqrt{V}\,e_a = \delta_a^\mu(\p_\mu+2MA_\mu\p_t)\,,
\end{split}
\ee
where $A_\mu$ are the spacetime components of the SD dyon $A=\d t/r+a$ in the Cartesian coordinates $x^\mu=(t,x^i)$. The second equality follows by substituting $V\p_t=(1+2M/r)\p_t$. 

Now, take the time dependence of $\psi_{\alpha\cdots\beta}$ to be a factor of  $\e^{\im\omega t}$. We can then replace the $\p_t$ by $\im\omega$ where desired.  Thus, when acting on $\psi_{\alpha\cdots\beta}$ we can write
\be
\sqrt{V}\,\nabla_a\psi_{\alpha\cdots\beta} = \sqrt{V}\,e_a\psi_{\alpha\cdots\beta} = \delta_a^\mu(\p_\mu+2MA_\mu\p_t)\psi_{\alpha\cdots\beta}\,.
\ee
After multiplying  through \eqref{SDTN-ZRM} by $\sqrt{V} $, the zero-rest-mass equation reduces to:
\begin{equation}
\left(\p^{\al\dal}+2\im M\omega A^{\al\dal}\right)\psi_{\alpha\alpha_2\ldots \alpha_n}=0\
\,.
\end{equation}
Here, $A_{\al\dal}$ are the components of the SD dyon in the flat space co-frame $\d x^{\al\dal}$. We see that \eqref{SDTN-ZRM} is reduced to the background-coupled ASD wave equation on a SDD for fields of charge $e=2M\omega$.

A similar procedure works for the scalar case $n=0$. We start with the Laplacian of a scalar $\vphi$,
\be
\Box\vphi = \frac{1}{|g|^{1/2}}\,\p_\mu(|g|^{1/2}g^{\mu\nu}\p_\nu\vphi)\,,
\ee
with $|g|$ the metric determinant, and try to express it in terms of the tetrad. Take the indices $\mu,\nu$ to again refer to the coordinates $x^\mu = (t,x^i)$. Using $g^{\mu\nu} = \delta^{ab}e_a^\mu e_b^\nu$, this becomes
\be
\begin{split}
    \Box\vphi &= \frac{1}{|g|^{1/2}}\,\p_\mu(|g|^{1/4}e_a^\mu\cdot|g|^{1/4} e^a\vphi)\\
    &= \frac{1}{|g|^{1/4}}\,e_a(|g|^{1/4}e^a\vphi) + \frac{1}{|g|^{1/4}}\,\p_\mu(|g|^{1/4}e_a^\mu)\,e^a\vphi\,,
\end{split}
\ee
where we have abbreviated $e^{a\nu}\p_\nu\vphi = e^a\vphi$, etc. Observe that the second term in this expression vanishes: using $|g|=V^2$, it follows that the vectors $|g|^{1/4}e_a$ are divergence-free
\be
\begin{split}
    \p_\mu(|g|^{1/4}e_a^\mu) &= \p_\mu\left(\delta_a^\mu + 2M\delta^\nu_a A_\nu\delta^\mu_t\right)\\
    &= 2M\,\delta^\nu_a\,\p_tA_\nu = 0\,,
\end{split}
\ee
as a consequence of the time-independence of the SDD $A_\mu$.

This means that the Laplacian on SDTN is simply
\be
\begin{split}
    \Box\vphi &= \frac{1}{\sqrt{V}}\,e_a\big(\sqrt{V}e^a\vphi\big)\\
    &= \frac{1}{V}\,\delta^{\mu\nu}(\p_{\mu}+2MA_{\mu}\p_t)(\p_{\nu}+2MA_{\nu}\p_t)\vphi\,.
\end{split}
\ee
If $\vphi$ is a momentum eigenstate carrying energy $\omega$, this reduces to
\be
\Box\vphi = \frac{1}{V}\,\delta^{\mu\nu}(\p_{\mu}+2M\im\omega A_{\mu})(\p_{\nu}+2M\im\omega A_{\nu})\vphi\,,
\ee
having again used the time-independence of $A_\nu$ to commute the $\p_t$ in the first bracket past the $A_\nu$ in the second. Up to the prefactor of $1/V$, this is precisely the gauge covariant Laplacian in the SD dyon background! 

We can therefore lift  our solutions from the previous subsection. Recall that we must have that $2e\in\Z$ to make $\psi_{\alpha\cdots\beta}$ (or the scalar $\vphi$) single-valued, so the energy $\omega$ is necessarily quantized to satisfy $4M\omega\in\Z$. Of course, this also follows on SDTN from the periodicity in $t$. That is to say, we can construct helicity $-n/2$ minimal states around SDTN as
\begin{empheq}[box=\fbox]{align}
\hspace{1em}\varphi_{\alpha_1\ldots\alpha_n}^+=\kappa_{\alpha_1}\ldots\kappa_{\alpha_n}\left(\frac{r}{1+|\zeta|^2}\right)^{2M\omega}(\bar\zeta z+1)^{4M\omega}\e^{\im k\cdot x}\,,\hspace{1em}
\end{empheq}
for positive energy, and
\begin{empheq}[box=\fbox]{align}
\hspace{1em}\varphi_{\alpha_1\ldots\alpha_n}^-=\kappa_{\alpha_1}\ldots\kappa_{\alpha_n}\left(\frac{r}{1+|\zeta|^2}\right)^{-2M\omega}(\zeta -z)^{-4M\omega}\e^{\im k\cdot x}\,,\hspace{1em}
\end{empheq}
for negative energy, respectively.

We reiterate  that in this section we did not use the special form of $V$ or $a$ for the SD Taub-NUT spacetime. Thus, this technique can be applied to \emph{any} Gibbons-Hawking metric to obtain background-coupled massless fields from charged fields with respect to the static SD Maxwell field $A_\mu =(V-1,a)$.

\paragraph{Potentials:}
In the special case of a negative-helicity graviton, we take $n=4$.  In order to construct a 2-point function, we will need not only the ASD part of the linearized Weyl tensor $\psi_{\alpha\beta\gamma\delta}$, but also the corresponding linear perturbation $\gamma_{\alpha\beta}$ to the ASD spin connection. Introducing the spinor equivalent $\theta^{\alpha\dot\alpha}$ of the coframe and the ASD 2-forms $\Sigma^{\alpha\beta}\coloneqq\theta^{\alpha}{}_{\dot\alpha}\wedge\theta^{\beta\dot\alpha}$, the linearized spin connection is a 1-form satisfying
\be
\d \gamma_{\alpha\beta}=\psi_{\alpha\beta\gamma\delta}\,\Sigma^{\gamma\delta}\,.
\ee
Its components in the chosen coframe,  $\gamma_{\alpha\beta}=\gamma_{\alpha\beta\delta\dot\delta}\theta^{\delta\dot\delta}$ can be expressed by introducing a pair of \emph{dressing matrices} whose form is motivated by the dressed momentum \eqref{eq:dressed_momenta_dyon}
\begin{align}
    H^{\pm\,\dal}{}_{\dot\beta}(x,\lambda)&=\frac{1}{\sqrt{V}}\left(\delta^{\dal}_{\dot\beta}-\frac{4M}{r}\frac{T^{\alpha\dot\alpha}\kappa_\alpha\,\chi_\pm^\beta\, T_{\beta\dot\beta}}{\langle \kappa\,\chi^\pm\rangle}\right)\,,\label{eq:pos_frame}
\end{align}
These matrices are unimodular and satisfy the relations
\be
K^{\pm}_{\dal}=\tilde\kappa_{\dot\beta}H^{\pm\,\dot\beta}{}_{\dal}\,,\qquad\d(\kappa_\alpha H^{\pm\,\dot\beta}{}_{\dal}\theta^{\alpha\dal})=0\,.\label{eq:frame_momentum_relation}
\ee
This directly implies that the potential
\begin{empheq}[box=\fbox]{align}
    \hspace{1em}\gamma^\pm_{\al\beta\gamma\dot\gamma} &= \frac{1}{[\eta\tilde\kappa]}\,\kappa_\al\kappa_\beta\kappa_\gamma\eta_{\dot\delta}H^{\pm\,\dot\delta}{}_{\dot\gamma}\vphi^\pm\,,\hspace{1em}\label{gammapm}
\end{empheq}
built from positive and negative energy scalar solutions $\vphi^\pm$ satisfies the linearized equations of motion for the ASD spin connection. Here $\eta_{\dot\alpha}$ is an arbitrary constant spinor which will drop out of gauge invariant observables.

%%%%%%%%%%%%%%%%%%%%%%

\subsection{Positive helicity states} 

In the remainder of this section, we describe positive helicity gluons on the SD dyon and positive helicity photons and gravitons on SDTN. These can be obtained by raising the helicity of the scalar solutions following \S6.4 of~\cite{Penrose:1986uia} or Plebanski-like formulations of the self-duality equations~\cite{Plebanski:1975wn}. This is most direct at the level of gauge potentials for the field strengths or curvature tensors.

For gluons, we seek linearized gauge fields with vanishing ASD linearized field strength. We can realize these states by linearizing the $K$-matrix formalism of self-dual Yang-Mills \cite{Mason:1991rf,Bogna:2023bbd}: introduce a scalar potential $\varphi$ for the gauge potential
\be
a_{\alpha\dot\alpha}=\xi_\alpha\,\xi^\beta\,\D_{\beta\dot\alpha}\vphi\,,\label{eq:pos-hel-gluon}
\ee
where $\xi_\alpha$ is an arbitrary reference spinor that will drop out from gauge invariant quantities, including scattering amplitudes. It follows from the self-duality condition of the background covariant derivative that the linearized ASD field strength is
\be
f_{\alpha\beta}=\frac{1}{2}\,\xi_\alpha\xi_\beta\,\D_{\gamma\dot\gamma}\D^{\gamma\dot\gamma}\vphi\,,\label{eq:f=0}
\ee
so that the state \eqref{eq:pos-hel-gluon} represents a positive-helicity gluon whenever $\varphi$ solves the charged scalar wave equation. Thus the operator $\xi_\alpha\xi^\beta\D_{\beta\dot\alpha}$ acts a helicity-raising operator. In terms of the dressed momenta, we can take the positive-helicity states to have the linearized gauge potential
\be
\boxed{\hspace{1em}a^\pm_{\alpha\dot\alpha}=\frac{1}{\langle \xi\kappa\rangle}\,\xi_\alpha \tilde{K}^\pm_{\dot\alpha}\vphi^\pm\hspace{1em}}
\ee
where factors of $\langle\xi\kappa\rangle$ have been introduced for normalization. 

\medskip

Next, we come to the SDTN case. Since the SDTN metric is self-dual as well as Ricci-flat, the spin connection acting on the bundle of left-handed Weyl spinors is flat. In the coframe \eqref{SDTN-ZRM}, a component-wise calculation shows that this left-handed spin connection vanishes identically. Consequently, in such a coframe, any constant spinor $\xi_\al$ on $\R^4$ is automatically covariantly constant with respect to the SD Taub-NUT metric. Hence, it can again be used to raise spin following the rules of~\cite{Hawking:1978ghb}.

On SDTN, the operator $\xi_\alpha\xi^\beta\nabla_{\beta\dot\alpha}$ still acts as a helicity-raising operator, where $\nabla_{\alpha\dot\alpha}$ are now the components of the covariant derivative in the basis $\theta^{\alpha\dot\alpha}$. This means that a positive-helicity photon is again described by the linearized gauge potential
\be\label{aph}
a_{\al\dal} = \xi_\al\, \xi^\beta\,\nabla_{\beta\dal}\vphi\,,
\ee
and that the linearized field strength of such a state is purely SD if $\varphi$ satisfies the wave equation. A positive-helicity spin-2 wavefunction can be constructed by applying the helicity-raising operator twice to the scalar solution:
\be\label{hph}
h_{\al\dal\beta\dot\beta} =  \xi_\al \xi_\beta \xi^\gamma \xi^\delta\nabla_{\gamma\dal}\nabla_{\delta\dot\beta}\varphi\,.
\ee
A calculation similar to the one leading to \eqref{eq:f=0} shows that the ASD part of its linearized Weyl tensor vanishes. This is a linearization of the formulation of self-dual Ricci-flat metrics in terms of Plebanski's second heavenly potential~\cite{Plebanski:1975wn}.

%%%%%%%%%%%%%%%%%%%%%%%
%%%%%%%%%%%%%%%%%%%%%%%

\section{Two-point scattering amplitudes}
\label{sec:amps}

In the perturbiner formalism, the tree-level two-point amplitude is captured by the on-shell kinetic term evaluated on a linear combination of external states in the background of choice. For scalar scattering around both backgrounds and for photon scattering around SD Taub-NUT, the usual non-chiral actions suffice. Conversely, it is natural to adopt chiral formulations of Yang-Mills~\cite{Chalmers:1996rq} and general relativity~\cite{Plebanski:1977zz} when computing gluon amplitudes around SD dyon and graviton amplitudes around SD Taub-NUT backgrounds. They express gauge theory and gravity as deformations around their self-dual sectors. 

These chiral formulations are built perturbatively around $BF$ type theories, where the linear field $B$ encodes the negative-helicity particles. The $BF$ theories capture the integrable self-dual sectors, so their scattering amplitudes vanish identically. As an important consequence, the two-point amplitudes we are interested in are zero for the $(+\,+)$ and $(+\,-)$ helicity configurations. This is because such amplitudes do not receive contributions from non-self-dual interactions. 

On the other hand, since the deformation away from self-duality is quadratic in the negative-helicity field, there is scope for non-vanishing $(-\,-)$ 2-point amplitudes. Calculating these is our main focus. We denote these amplitudes by $\cA(1_{e/\omega=\pm}^{h_1},2^{h_2}_{e/\omega=\mp})$, where $h_{1,2}$ denotes the helicity of each particle and the subscripts denote whether each particle has positive or negative charge/energy. Other selection rules also arise from the symmetries of the problem. The translation symmetry in $t$ imposes that 2-point amplitudes can only be non-vanishing if one of the states has positive energy and the other state has negative energy. Similarly, in the SD dyon case, the two states will be required to have equal and opposite charges due to global charge conservation. In what follows, we briefly touch upon each of the requisite chiral formulations, before using them to compute the desired amplitudes.

%%%%%%%%%%%%%%%%%%%%%%%
%%%%%%%%%%%%%%%%%%%%%%%
\subsection{Scalar amplitudes}

The free scalar action is 
\be
S_\text{scalar}=\int_M\d^4x\,\sqrt{|g|}\,g^{\mu\nu}\nabla_\mu\varphi\,\nabla_\nu\varphi\,.
\ee 
As advertised, we consider two cases: one in which $g$ is the flat metric and $\nabla$ is the covariant derivative $\D$ coupled to a Cartan-valued SDD embedded in some gauge algebra, and another in which $g$ is the SDTN metric and $\nabla$ the associated covariant derivative. 

Using \eqref{eq:dressed_momenta_dyon}, 
the scalar 2-point amplitude around both backgrounds is given by
\begin{multline}
\mathcal{A}(1^0_{e=-},2^0_{e=+})=2\pi\langle 1\,2\rangle \, \tilde \delta \\\times\int\d^3\vec x\left(\frac{r(\zeta-z_1)(\bar\zeta z_2+1)}{1+|\zeta|^2}\right)^n\left(\tilde\kappa_{1\,\dot\alpha}+\frac{n}{r}\frac{\chi^-_\alpha \tensor{T}{^\alpha_{\dot\alpha}}}{\langle 1\chi^-\rangle}\right)\left(\tilde\kappa_2^{\dot\alpha}+\frac{n}{r}\frac{\chi^+_\beta T^{\beta\dot\alpha}}{\langle 2\chi^+\rangle}\right)\e^{\im(\vec k_1+\vec k_2)\cdot\vec x}\,.
\end{multline}
The value of $n$ is $2e_2$ on the SDD and $4M\omega_2$ on SDTN, while $\tilde\delta$ equals $\delta_{-e_1,e_2}\delta(\omega_1+\omega_2)$ around SDD and equals $\delta_{-\omega_1,\omega_2}$ around SDTN. 

As we show in appendix \ref{app:int}, the integral can be evaluated explicitly and vanishes on the support of charge or energy conservation in the SD dyon and SD Taub-NUT cases respectively. That is,
\be
\boxed{\hspace{1em}\mathcal{A}(1^0_{e=-},2^0_{e=+})=0\,.\hspace{1em}}
\ee
This is consistent with a variety of observations. Its vanishing turns out to be closely related to gauge invariance of the 2-graviton $(-\,-)$ amplitude, see below. It is also expected from the fact that the free scalar action uplifts to a local holomorphic action on twistor space, and the spacetime states lift to twistor space states that do not scatter. The latter is essentially a consequence of the geometric fact that twistor space contains multiple copies of spacetime, and the two scalar states live on disjoint copies which do not interact \cite{Costello:2022wso}.

%%%%%%%%%%%%%%%%%%%%%%

\subsection{Photon scattering on SD Taub-NUT}

The standard QED action on a general curved spacetime can be used to compute the 2-point $(-\,-)$ photon amplitude on SDTN. In terms of the states previously introduced, the on-shell QED action reads
\be
\begin{aligned}
S_\text{QED}&=\frac{1}{4\tg^2}\int_MF_{1^-}\wedge F_{2^+}\\&=\int_M\d^4x\,V\,b^{-}_{1\,\alpha\beta}\,b^{+\,\alpha\beta}_{2}\,,
\end{aligned}
\ee
where $\tg$ is the Maxwell coupling, $b^{-}_{1\,\alpha\beta}$ is given by \eqref{eq:phi-} with $n=2$ and $e=2M\omega_1<0$, and $b^{+}_{2\,\alpha\beta}$ is given by \eqref{eq:phi+} with $n=2$ and $e=2M\omega_2>0$. In particular, we have taken photon 1 to have negative energy and photon 2 to have positive energy, and used the ASD condition on the field strength. Upon inserting the expressions for the quasi-momentum eigenstates and performing the integral along the time direction, the 2-point amplitude reads
\begin{multline}
\mathcal{A}(1^{-1}_{\omega=-},2^{-1}_{\omega=+})=2\pi\,\langle 1\,2\rangle^2\,\delta_{-\omega_1,\omega_2}\\\times\int\d^3\vec x\,\left(1+\frac{2M}{r}\right)\left(\frac{r(\zeta-z_1)(\bar\zeta z_2+1)}{1+|\zeta|^2}\right)^{4M\omega_2}\e^{\im(\vec k_1+\vec k_2)\cdot x}\,.
\end{multline}
This integral can be evaluated by means of the techniques developed in appendix \ref{app:int} and it vanishes:
\be
\boxed{\hspace{1em}
\mathcal{A}(1^{-1}_{\omega=-},2^{-1}_{\omega=+})=0\,.
\hspace{1em}}
\ee

%%%%%%%%%%%%%%

\subsection{Gluon scattering on SD dyon}
\label{sec:dyonamps}

For sake of simplicity, let us embed the SD dyon in an $\SU(2)$ gauge theory by taking the Cartan-valued color factor in \eqref{sdc} to be $\sc=\sigma_3$, the third Pauli matrix. Then the off-diagonal gluons have positive and negative charge. 

The relevant chiral action for gauge theory is due to Chalmers and Siegel~\cite{Chalmers:1996rq},
\be
S[B,A]=\int \tr\, B\wedge F-\frac{\tg^2}{2}\int \tr\, B\wedge B\,.
\ee
Here, $B$ is an adjoint-valued ASD 2-form, $F$ is the field strength of the gauge field $A$ and $\tg$ is the Yang-Mills coupling. The equation of motion for $B$ ensures that the ASD component of $F$ is proportional to $B$ itself, so that the on-shell action reduces to
\begin{equation}
    S_{\text{on-shell}}[B,A]=\frac{\tg^2}{2}\int\tr\, B\wedge B\,.
\end{equation}
By virtue of being ASD, $B$ can be expanded as $B=B_{\al\beta}\Sigma^{\al\beta}$ on a basis of ASD 2-forms $\Sigma^{\al\beta}$.\footnote{Recall that in flat space, we are working with $\Sigma^{\al\beta} = \d x^{\al}{}_{\dal}\wedge\d x^{\beta\dot\alpha}$.} The linearized equation of motion of $F$ becomes the helicity $-1$ free field equation for $B_{\al\beta}$. Therefore $B$ gets identified with negative helicity gluons.

As previously discussed, the two-point $(-\,-)$ tree-level amplitude is constructed by inserting the field $B=\tg^{-1}(b_1+b_2)$, where the $b_i$ are negative helicity quasi-momentum eigenstates. The trace over the colour indices implies that the amplitude can be non-vanishing only if the charge is conserved, $e_1+e_2=0$. And as usual we have energy conservation $\omega_1+\omega_2=0$ as a consequence of the time translation symmetry. 

We will assume without loss of generality that $b_1$ has a negative charge $e_1<0$. After performing the integral over $t$, the two-point amplitude reduces to
\begin{multline}
   \mathcal{A}(1^{-1}_{e=-},2^{-1}_{e=+})=-\pi\, \langle 1\,2\rangle^2\, \delta(\omega_1+\omega_2)\,\delta_{-e_1,e_2}\\\times\int\d^3\vec x\,\left(\frac{r(\zeta-z_1)(\bar\zeta z_2+1)}{1+|\zeta|^2}\right)^{2e_2}\e^{\im(\vec k_1+\vec k_2)\cdot\vec x}\,.\label{eq:integral-2-point-gluon}
\end{multline}
Since the background has a time-like, as well as a spatial $\SO(3)$, symmetry, we expect the 2-point amplitude to possess these symmetries as well. The time translation symmetry is manifest from the energy-conserving delta function, while the $\SO(3)$ invariance can be made manifest by explicitly computing the integral in \eqref{eq:integral-2-point-gluon}. We defer the detailed computation of such integrals to appendix \ref{app:int}, stating the final expression here:
\be
\boxed{\hspace{1em}\mathcal{A}(1^{-1}_{e=-},2^{-1}_{e=+})=- 8\pi^2\,e_2\,(2e_2)!\,\frac{(2\omega_1)^{2e_2-1}\,\la 1\,2\ra^{2+2e_2}}{|\vec{k}_1+\vec{k}_2|^{4e_2+2}\, (\kappa_{1}^0\,\kappa_{2}^0)^{2e_2}}\,\delta(\omega_1+\omega_2)\,\delta_{-e_1,e_2}\,.\hspace{1em}}\label{eq:integrated-2-point-gluon}
\ee
The factor of $(\kappa_{1}^0\,\kappa_{2}^0)^{-2e_2}$ ensure that the wave functions, and hence the amplitude, has the appropriate little group scaling: as discussed, this needs to be corrected as a consequence of the non-trivial topology of the SDD background. Rotational invariance can be manifested by dropping the factor of $(\kappa_{1}^0\,\kappa_{2}^0)^{-2e_2}$ from the wave functions at the price of introducing an anomalous little-group weight. 

%%%%%%%%%%%%%%
%%%%%%%%%%%%%%

\subsection{Graviton scattering on SD Taub-NUT}
\label{sec:tnamps}

General relativity admits a first-order chiral formulation, first developed by Plebanski~\cite{Plebanski:1977zz}; see~\cite{Krasnov:2009pu} for a review. The field variables are the tetrad $\theta^{\alpha\dot\alpha}$ and an ASD spin connection $\Gamma_{\alpha\beta}=\Gamma_{(\alpha\beta)}$; the classical action is
\be
S[\theta,\Gamma]=\int_M\Sigma^{\alpha\beta}\wedge\d\Gamma_{\alpha\beta}+\kappa^2\int_M\Sigma^{\alpha\beta}\wedge\tensor{\Gamma}{_\alpha^\gamma}\wedge\tensor{\Gamma}{_{\beta\gamma}}\,.
\ee
Here $\Sigma^{\alpha\beta}=\epsilon_{\dot\alpha\dot\beta}\theta^{\alpha\dot\alpha}\wedge\theta^{\beta\dot\beta}$, while $\kappa^2=16\pi G$ is the gravitational coupling. The classical equation of motion for $\Gamma_{\alpha\beta}$ sets $\kappa^2\Gamma_{\alpha\beta}$ equal to the ASD spin connection of the metric $g=\epsilon_{\alpha\beta}\epsilon_{\dot\alpha\dot\beta}\theta^{\alpha\dot\alpha}\theta^{\beta\dot\beta}$ on $M$. On the support of this equation of motion, the remaining equation of motion for $\Sigma^{\alpha\beta}$ reproduces Einstein's vacuum equations, so that the on-shell action is vanishing. 

However, if we consider a linear perturbation to the spin connection of the form $\delta\Gamma^{\alpha\beta}=\kappa^{-1}(\gamma_1^{\alpha\beta}+\gamma_2^{\alpha\beta})$ and identify the perturbations with negative-helicity quasi-momentum graviton eigenstates, the on-shell action is non-vanishing and computes the two-point $(-\,-)$ amplitude around the spacetime $M$ with background, non-dynamical metric $g$. Thus, the two-point amplitude is compactly given by
\be
\mathcal{A}(1^{-2},2^{-2})=\int_M\Sigma^{\alpha\beta}\wedge\tensor{\gamma}{_{1\,\alpha}^\gamma}\wedge\gamma_{2\,\beta\gamma}\,.
\ee
Upon inserting the on-shell states \eqref{gammapm} and assuming without loss of generality that graviton 1 has negative energy, the 2-point amplitude reads
\begin{multline}
    \mathcal{A}(1^{-2}_{\omega=-},2^{-2}_{\omega=+}) = -\frac{2\pi\,\la 1\,2\ra^3\,\eta_{1\,\dal}\,\eta_{2\,\dot\beta}}{[\eta_1\,1][\eta_2\,2]}\:\delta_{-\omega_1,\omega_2}\\
    \times\int\d^3\vec{x}\;V\,H_1^{-\,\dal}{}_{\dot\gamma}H_2^{+\,\dot\beta\dot\gamma}\left(\frac{r(\zeta-z_1)(\bar\zeta z_2+1)}{1+|\zeta|^2}\right)^{4M\omega_2}\e^{\im(\vec{k}_1+\vec k_2)\cdot\vec{x}}\label{eq:integral-2-point-graviton}\,,
\end{multline}
where $\eta_{1,2}$ are reference spinors parametrizing the residual gauge symmetry arising from linearized spin-frame variations and diffeomorphisms. The 2-point amplitude should be independent of the choice of these spinors. 

As previously anticipated, this gauge invariance is also closely related to the vanishing of the scalar amplitude around SDTN. Recalling the first equation \eqref{eq:frame_momentum_relation} relating the background-dressed momenta and the dressing matrices, we can express the scalar and graviton amplitudes as
\begin{align}
   \mathcal{A}(1^{0}_{\omega=-},2^{0}_{\omega=+})&\propto\tilde\kappa_{1\,\dot\alpha}\tilde\kappa_{2\,\dot\beta}\int\d^3\vec x\,VH_1^{-\,\dal}{}_{\dot\gamma}H_2^{+\,\dot\beta\dot\gamma}\left(\frac{r(\zeta-z_1)(\bar\zeta z_2+1)}{1+|\zeta|^2}\right)^{4M\omega_2}\e^{\im(\vec k_1+\vec k_2)\cdot\vec x}\,,\\
    \mathcal{A}(1^{-2}_{\omega=-},2^{-2}_{\omega=+})&\propto\frac{\eta_{1\,\dot\alpha}\eta_{2\,\dot\beta}}{[\eta_1\,1][\eta_2\,2]}\int\d^3\vec x\,VH_1^{-\,\dal}{}_{\dot\gamma}H_2^{+\,\dot\beta\dot\gamma}\left(\frac{r(\zeta-z_1)(\bar\zeta z_2+1)}{1+|\zeta|^2}\right)^{4M\omega_2}\e^{\im(\vec k_1+\vec k_2)\cdot\vec x}
\end{align}
up to kinematical prefactors. Gauge invariance of the graviton 2-point amplitude requires the integral to be proportional to $\tilde\lambda_1^{\dot\alpha}\tilde\lambda_2^{\dot\beta}$, and this in turns implies the vanishing of the scalar 2-point amplitude.

Finally, the evaluation of the integral in \eqref{eq:integral-2-point-graviton} leads to the following expression for the 2-point amplitude
\be
\boxed{\hspace{1em}\mathcal{A}(1^{-2}_{\omega=-},2^{-2}_{\omega=+}) = 4\pi^2\,M\,(4M\omega_2)!\,\frac{(2\omega_1)^{4M\omega_2-2}\,\la 1\,2\ra^{4+4M\omega_2}}{|\vec{k}_1+\vec{k}_2|^{8M\omega_2+2}\, (\kappa_{1}^0\,\kappa_{2}^0)^{4M\omega_2}}\,\delta_{-\omega_1,\omega_2}\,,\hspace{1em}}\label{eq:integrated-2-point-graviton}
\ee
Again the factor of $(\kappa_{1}^0\,\kappa_{2}^0)^{-4M\omega_2}$ comes from the original wave functions and can be dropped to manifest rotational invariance at the price of introducing an anomalous little-group weight. 
To obtain the corresponding formula \eqref{eq:integrated-2-point-graviton-spin} for the spinning SD Taub-NUT, we must perform the imaginary shift along $i\vec{a}$ to reduce the centre of the spinning SD Taub-NUT to $\vec{x}=0$ as discussed following \eqref{SDTN-KS}.  This leads to the integral \eqref{eq:integral-2-point-graviton} above except that the factors of $\e^{i\vec{k}_i\cdot x}$ generate the extra exponential factor of $\e^{-(\vec{k}_1+\vec{k}_2)\cdot\vec{a}}$  appearing in \eqref{eq:integrated-2-point-graviton-spin} under this shift.

%%%%%%%%%%%%%%%%%%%%%%%
%%%%%%%%%%%%%%%%%%%%%%%

\subsection{Double copy structure}
\label{sec:double}

Double copy refers to a broad set of methods for writing perturbative scattering amplitudes of gravitational theories as the `square' of perturbative amplitudes of some non-gravitational theories (cf., \cite{Bern:2019prr,Borsten:2020bgv,Adamo:2022dcm} for reviews). In its most developed form, this relates multiparticle graviton scattering amplitudes of general relativity to gluon scattering amplitudes of Yang-Mills theory, and there are hints that some of that correspondence continues to hold in curved backgrounds~\cite{Adamo:2017nia,Adamo:2018mpq,Adamo:2020qru}. Double copy has also been extended to algebraically special classes of exact solutions to give a notion of `classical double copy'~\cite{Monteiro:2014cda}. The SD Taub-NUT metric is the classical double copy of a SD dyon in this language~\cite{Luna:2015paa}, and we have seen the double copy at work in the construction of background coupled fields. 
In particular, squaring the non-exponential part of a spin-1 background coupled field on the SDD gives a spin-2 background coupled field on SDTN under $e\to M\omega$. This only respects the quantization condition for $e$ even because conventionally we do not double the $\exp(\im k\cdot x)$ factor.  

One does not usually speak of double copy for 2-point amplitudes: in flat backgrounds these amplitudes are vanishing, and because they are determined by only the kinetic part of the theory they are insensitive to colour-kinematics duality. Nevertheless, as we have non-vanishing 2-point amplitudes in the $(-\,-)$ helicity configuration, it is natural to investigate the existence of any double copy relationship between \eqref{eq:integrated-2-point-gluon} and \eqref{eq:integrated-2-point-graviton}.

Clearly, upon squaring the 2-point gluon amplitude on the SDD a replacement rule is needed to remove the charges $e_{1,2}$. Since the 2-point amplitude is entirely governed by the inner product between free fields, via the kinetic term in the classical action, it seems most natural to implement the same replacement rule which takes charged zero-rest-mass fields on SDD to zero-rest-mass fields with the doubling described above, namely $e\to M\omega$. This leads to a simple double copy-like relation between the amplitudes:
\be\label{ampdc}
\mathcal{A}(1^{-2}_{\omega=-},2^{-2}_{\omega=+}) \sim \,|\vec{k}_ 1+\vec{k}_2|^2\,\mathcal{A}^2(1^{-1}_{e=-},1^{-1}_{e=+})\Big|_{e_{1,2}\to M\omega_{1,2}}\,,
\ee
where the amplitudes are implicitly understood to be stripped of charge/momentum conserving delta functions or Kronecker deltas, combinatorial factors etc.. At present, we lack any sort of first-principles understanding of this relationship.

%%%%%%%%%%%%%%%%%%%%%%%
%%%%%%%%%%%%%%%%%%%%%%%

\section{Interpretations and future directions}
\label{sec:conclusions}

We have shown that massless free field equations on self-dual dyon and Taub-NUT backgrounds can be solved with particularly simple quasi-momentum eigenstates. These are much simpler than solutions on the corresponding non-chiral Coulomb and Schwarzschild backgrounds (or generic dyon or Taub-NUT backgrounds), which are only expressible in harmonic expansions and in terms of special functions. This reflects the complete integrability underlying the SDD and SDTN backgrounds and allows us to perform the integrations explicitly to yield two-point functions. 

In the weak background field limit, these should correspond to a four point amplitude in a probe limit, i.e., with an incoming and outgoing massive leg whose momentum is unchanged, interacting with an incoming and subsequently outgoing massless probe particle. However, the formuale in~\cite{Johansson:2019dnu,Aoude:2020onz} (for instance), assume momentum conservation, whereas our integration procedure has focused on the part of the two-point function that is not supported on conserved momentum. Thus, implementing the perturbative limit at the level of the final 2-point amplitudes is probably not the correct way to proceed. Instead, one could implement the perturbative limit \emph{before} performing any integrals, by linearising the integrand of \eqref{eq:integral-2-point-graviton} in $M$ (in the gravitational case) and then integrating. Indeed, this is how the perturbative limit is implemented in other contexts of scattering on other backgrounds (cf., \cite{Adamo:2020yzi,Adamo:2020qru,Adamo:2021hno,Fedotov:2022ely,Adamo:2022mev,Adamo:2023cfp}), where translation invariance is also broken. We hope to address this point in future work. 

More generally, our integrated results are closer to describing an amplitude for a negative helicity massless particle of momentum $\vec{k_1}$ having its helicity flipped by the SD background into one of positive helicity with momentum $\vec{k_2}$, as described for general MHV amplitudes in~\cite{Mason:2008jy}. It would be interesting to see if some correspondence can be established with the integrable classical particle scattering results of~\cite{Dunajski:2019gwz}.

Our backgrounds do have the feature of non-trivial topology, leading to a non-trivial spin weight for charged probe particles on the SDD, and particles with (quantized) non-zero energy on SDTN.  This in turn leads to higher spherical harmonics and corresponding higher growth in $r$ than expected in flat space. It would be interesting to recover these results from previous investigations of dyon scattering, based on harmonic expansions, by restricting to the SD sector and demanding regularity on the sphere~\cite{Schwinger:1976fr,Boulware:1976tv}.  

We introduced spin into our two-point functions in \eqref{eq:integrated-2-point-graviton-spin} following  the Newman-Janis correspondence~\cite{Newman:1965tw,Huang:2019cja,Emond:2020lwi}, which in the self-dual case becomes a simple imaginary shift; with this we obtain spinning SD Taub-NUT as a complex metric  providing the self-dual part of the Kerr metric. This imaginary shift in the direction of the spin vector $\vec{a}$ can be used to move  the contour of integration so that the integrand in the spinning case is the same as given above as in for example \eqref{eq:integral-2-point-gluon} or \eqref{eq:integral-2-point-graviton}, except that the plane wave factors in the wave function must be shifted to give the additional factor of $\e^{-(\vec{k_1}+\vec{k_2})\cdot \vec{a}}$.  Once this factor is taken outside the integrals, the rest reduces to those considered above leading to the same integrated answer \eqref{eq:integrated-2-point-gluon} or \eqref{eq:integrated-2-point-graviton}, but now with the overall additional factor of $\e^{-(\vec{k_1}+\vec{k_2})\cdot \vec{a}}$.  This is a general phenomenon and explains the simple exponential dependence on the spin found in~\cite{Arkani-Hamed:2017jhn,Aoude:2020onz} and implicitly in~\cite{Johansson:2019dnu}.  These relate to formulae with just ASD particles on the full background because the two ASD particles on the background will not interact with the ASD  part of the background to the orders considered.

For Coulomb solutions or physical black holes (Schwarzschild or Kerr) in Lorentzian signature, the topological issues that we have confronted here, of course, do not arise. This makes our results connected more directly to backgrounds with a non-vanishing magnetic charge or NUT parameter. However, it also makes it clear that if one wishes to pursue the strategy of perturbatively reconstructing the non-chiral background by expanding around the SD sector, the topological issues that we have encountered will arise at the first step.   
According to the Newman-Janis construction,  the Kerr metric is expressed as a nonlinear superposition of  SD Taub-NUT and its conjugate ASD Taub-NUT, displaced along the imaginary spin vector $\pm \im\,\vec{a}$.  In this  way, that the topology of the SDTN is canceled by its conjugate.  The wire singularity then needs only connect the two complex conjugate centres, reminiscent of the Kerr worldsheet of~\cite{Guevara:2020xjx}.

Our results relied on two novel techniques. The first was the introduction of charged Killing spinors that convert flat space massless fields to those that are charged with respect to the dyon background but with a different spin.  The second was the lifting of charged massless fields on the SD dyon background to SD Taub-NUT.  It will be interesting to explore the full domain of applicability of these techniques.  We have already remarked that the lifting described in \S\ref{SDTN-KS} of massless fields  that are background coupled to a self-dual static Maxwell  field to those that are coupled to the corresponding Gibbons-Hawking metric. Thus this will apply also to the ALE or ALF $A_k$ gravitational instantons. One can show via twistor methods that many self-dual Maxwell fields that admit charged Killing spinors can be constructed from twistor space, not just the SD dyon.  It would be interesting to see whether they can be extended beyond these classes, for example the metrics of \cite{Tod:1983pm} looking promising.

Non-trivial SD solutions are never real in Lorentzian signature, but can be in Euclidean or split signature and indeed the original SDTN was conceived as a Euclidean signature instanton~\cite{Hawking:1976jb}. When expanding around the SD sector to obtain results relevant to Lorentzian signature, one  envisions representing the SD background as a complex solution defined on what will eventually be the physical, Lorentzian-real slice\footnote{This perspective was first developed in~\cite{HawkingYM,Hawking:1979pi} for the computation os scattering on instantons via Wick rotation.}. However, it may then be reasonable to hope that the contours for the various integrations be deformable to other signatures, particularly when Feynman propagators are used as these naturally continue to the complex.  

On the Euclidean signature background, the full spheres of constant radius $r$ are the same as those for Lorentzian signature, and so the consequences of non-trivial topology cannot be avoided. However, for split (or neutral) signature, these spheres are replaced by hyperbolic discs, as in~\cite{Crawley:2021auj}, (or $1+1$ de Sitter) and so the topological issues are no longer present. It is conceivable that such choices better reflect an effective strategy for perturbing around the SD sector. As a possible application, the states \eqref{eq:general_wavefunctions} are non-singular even when either $p$ or $q$ are negative, because the on-shell condition in split signature implies that $z_i$ lies on the boundary of the unit disc in the complex plane, while $|\zeta|\leq 1$. In particular, this means  that we can scatter states with $n=0$, that is states with the standard $O(r^0)\,\e^{\im k\cdot x}$ behaviour at infinity. Furthermore, if Lorentzian momenta are chosen, the poles below can be taken to be in the complement of the disc over which the integration  takes place.  For example, the gluon amplitude on the SDD background in split signature takes  the form \begin{multline}
\mathcal{A}(1^{-1}_{e=-},2^{-1}_{e=+})=-\pi\,\delta(\omega_1+\omega_2)\,\delta_{-e_1,e_2}\,\langle 1\,2\rangle^2\\ \times\int_0^\infty\d r\,r^2\int_{\HH_2}\frac{\d \zeta\,\d\bar\zeta}{(1-\zeta\bar\zeta)^2}\left[\frac{(\zeta -z_1)(\bar\zeta z_2-1)}{(\zeta-z_2)(\bar\zeta z_1-1)}\right]^{e_2}\e^{\im(\vec k_1+\vec k_2)\cdot\vec x}\,.
\end{multline}
Here $(\zeta,\bar\zeta)$ are Poincar\'e disc coordinates on the hyperbolic space $\HH_2$. If we take the momenta in split signature, $|z_i|=1$ whereas in Lorentz signature we can take  $|z_1|<1$ and $|z_2|>1$. We leave a systematic evaluation of the integrals in these amplitudes to future work.

%%%%%%%%%%%%%%%%%%%%%%%
%%%%%%%%%%%%%%%%%%%%%%%

%%%%%%%%%%%%%%%%%%%%%%%
%%%%%%%%%%%%%%%%%%%%%%%

\acknowledgments 
We are grateful to Roland Bittleston, Erin Crawley, Alfredo Guevara, Elizabeth Himwich, Adam Kmec, Uri Kol, Alex Ochirov, Donal O'Connell, Natalie Paquette, Martin Ro\v{c}ek, Andrew Strominger and Chiara Toldo for helpful discussions. TA is supported by a Royal Society University Research Fellowship, the Leverhulme Trust grant RPG-2020-386 and the Simons Collaboration on Celestial Holography MP-SCMPS-00001550-11. GB is supported by a joint Clarendon Fund and Merton College Mathematics Scholarship. LJM would like to thank the Institutes des Haut \'Etudes Sci\'entifique, Bures Sur Yvette, and the Laboratoire Physique at the ENS, Paris for hospitality while this was being written up and the STFC for financial support from  grant number ST/T000864/1. AS is supported by a Black Hole Initiative fellowship, funded by the Gordon and Betty Moore Foundation and the John Templeton Foundation. 

%%%%%%%%%%%%%%%%%%%%%%%
%%%%%%%%%%%%%%%%%%%%%%%

\appendix

%%%%%%%%%%%%%%%%%%%%%%%
%%%%%%%%%%%%%%%%%%%%%%%

\section{Integration of the two-point amplitudes}
\label{app:int}
Here we give some details on the integration of the two-point amplitudes. These will be made up from  the six basis integrals
\begingroup
\allowdisplaybreaks
\begin{align}
\mathscr{H}_n(\vec k,z_1,z_2)&=\int\d^3\vec x\,\left(\frac{r(\zeta-z_1)(\bar\zeta z_2+1)}{1+|\zeta|^2}\right)^{n}\e^{\im \vec k\cdot\vec x}\,,\label{eq:In-integral}\\
\mathscr{J}_n^0(\vec k,z_1,z_2)&=\int\d^3\vec x\,\left(\frac{r(\zeta-z_1)(\bar\zeta z_2+1)}{1+|\zeta|^2}\right)^{n-1}\frac{1}{1+|\zeta|^2}\,\e^{\im \vec k\cdot\vec x}\,,\label{eq:J0n-integral}\\
\mathscr{J}_n^\zeta(\vec k,z_1,z_2)&=\int\d^3\vec x\,\left(\frac{r(\zeta-z_1)(\bar\zeta z_2+1)}{1+|\zeta|^2}\right)^{n-1}\frac{\zeta}{1+|\zeta|^2}\,\e^{\im \vec k\cdot\vec x}\,,\label{eq:Jzetan-integral}\\
\mathscr{J}_n^{\bar\zeta}(\vec k,z_1,z_2)&=\int\d^3\vec x\,\left(\frac{r(\zeta-z_1)(\bar\zeta z_2+1)}{1+|\zeta|^2}\right)^{n-1}\frac{\bar\zeta}{1+|\zeta|^2}\,\e^{\im \vec k\cdot\vec x}\,,\label{eq:Jbarzetan-integral}\\
\mathscr{J}_n^{\zeta\bar\zeta}(\vec k,z_1,z_2)&=\int\d^3\vec x\,\left(\frac{r(\zeta-z_1)(\bar\zeta z_2+1)}{1+|\zeta|^2}\right)^{n-1}\frac{|\zeta|^2}{1+|\zeta|^2}\,\e^{\im \vec k\cdot\vec x}\,,\label{eq:Jzetabarzetan-integral}\\
\mathscr{K}_n(\vec k,z_1,z_2)&=\int\d^3\vec x\,\left(\frac{r(\zeta-z_1)(\bar\zeta z_2+1)}{1+|\zeta|^2}\right)^{n-1}\frac{1}{r}\,\e^{\im \vec k\cdot\vec x}\,,\label{eq:Kn-integral}
\end{align}
\endgroup
where $n\in\Z_{\geq0}$ is a non-negative integer.  Here $\vec{k}=\vec{k}_1+\vec{k}_2$ with $\vec{k}_1,\vec{k}_2$ the momenta of of the two wave functions. We will compute such integrals assuming $\vec{k}$ to be real and analytically continue the final results to complex $\vec{k}$. This is equivalent to Wick rotating $t\mapsto\im t$ to turn the SDTN metric into a complex metric on a real slice on which the scattering particles have Lorentzian momenta. This follows the   computation of scattering amplitudes on instantons via Wick rotation, first developed in~\cite{HawkingYM,Hawking:1979pi}.

These are oscillatory integrals so we need to use an $\im\eps$ prescription to make the integrals convergent. Thus  we introduce a Gaussian factor of $\e^{-\eps(|\zeta|^2+r)}$ into the integrand, and take the limit $\eps\to0$. We will ignore the distributional terms supported on $|\vec k|=0$ that will also be present which correspond to forward scattering contributions.   
We discard these taking merely the \emph{na\"ive} $\eps\to0$ limit.

\subsection{Decomposition of the amplitudes}
In terms of this basis of integrals, the scalar amplitude can be expressed as
\begingroup
\allowdisplaybreaks
\begin{align}
    \mathcal{A}(1^0_{e=-},2^0_{e=+})&=2\pi\delta_{-\omega_1,\omega_2}\langle 1\,2\rangle\left[-[12]\mathscr{H}_n(\vec k_1+\vec k_2,z_1,z_2)-\frac{1}{2}n^2\mathscr{K}_n(\vec k_1+\vec k_2,z_1,z_2)\right.\nonumber\\&\quad -n(\iota_\alpha T^{\alpha\dot\alpha}\tilde\kappa_{2\,\dot\alpha}+z_1o_\alpha T^{\alpha\dot\alpha}\tilde\kappa_{1\,\dot\alpha})\mathscr{J}_n^0(\vec k_1+\vec k_2,z_1,z_2)\nonumber\\&\quad+(o_\alpha T^{\alpha\dot\alpha}\tilde\kappa_{1\,\dot\alpha}+o_\alpha T^{\alpha\dot\alpha}\tilde\kappa_{2\,\dot\alpha})\mathscr{J}_n^\zeta(\vec k_1+\vec k_2,z_1,z_2)\nonumber\\&\quad-(z_1\iota_\alpha T^{\alpha\dot\alpha}\tilde\kappa_{1\,\dot\alpha}+z_2\iota_\alpha T^{\alpha\dot\alpha}\tilde\kappa_{2\,\dot\alpha})\mathscr{J}_n^{\bar\zeta}(\vec k_1+\vec k_2,z_1,z_2)\nonumber\\&\left.\quad+(z_2o_\alpha T^{\alpha\dot\alpha}\tilde\kappa_{2\,\dot\alpha}+\iota_\alpha T^{\alpha\dot\alpha}\tilde\kappa_{1\,\dot\alpha})\mathscr{J}_n^{\zeta\bar\zeta}(\vec k_1+\vec k_2,z_1,z_2)\right]\,,
\end{align}
\endgroup
with $n=2e_2$ for scattering around SD dyon and $n=4M\omega_2$ for scattering around SD Taub-NUT, respectively. We have also introduced the constant spinors $\iota^\alpha\coloneqq(1,0)$ and $o^\alpha\coloneqq(0,-1)$, forming a dyad normalized as $\langle \iota o\rangle=1$. Similarly, the photon amplitude around SD Taub-NUT is
\be
\mathcal{A}(1^{-1}_{\omega=-},2^{-1}_{\omega=+})=2\pi\langle 1\,2\rangle^2\delta_{-\omega_1,\omega_2}(\mathscr{H}_{4M\omega_2}(\vec k_1+\vec k_2,z_1,z_2)+2M\mathscr{K}_{1+4M\omega_2}(\vec k_1+\vec k_2,z_1,z_2))\,.
\ee
The 2-point gluon amplitude can be expressed as
\be
\mathcal{A}(1^{-1}_{e=-},2^{-1}_{e=+})=-\pi \delta(\omega_1+\omega_2)\delta_{-e_1,e_2}\langle 1\,2\rangle^2\mathscr{H}_{2e_2}(\vec k_1+\vec k_2,z_1,z_2)\,,
\ee
with $n=2e_2$. Finally, upon expanding the product of the dressing matrices in \eqref{eq:integral-2-point-gluon} using \eqref{eq:pos_frame}, the 2-point graviton amplitude can be expressed in terms of these integrals as
\begingroup
\allowdisplaybreaks
\begin{align}
\mathcal{A}(1^{-2}_{\omega=-},2^{-2}_{\omega=+})&=-\frac{\langle 1\,2\rangle^3\eta_{1\,\dot\alpha}\eta_{2\,\dot\beta}}{[\eta_1\,1][\eta_2\,2]}2\pi\delta_{-\omega_1,\omega_2}\left(\varepsilon^{\dot\alpha\dot\beta}\mathscr{H}_{4M\omega_2}(\vec 
k_1+\vec k_2,z_1,z_2)\right.\nonumber\\&\quad+4MT^{\alpha\dot\alpha}T^{\beta\dot\beta}(\kappa_{1\,\alpha}\iota_\beta-z_1o_\alpha\kappa_{2\,\beta})\mathscr{J}^0_{4M\omega_2}(\vec k_1+\vec k_2,z_1,z_2)\nonumber\\&\quad+4MT^{\alpha\dot\alpha}T^{\beta\dot\beta}(-\kappa_{1\,\alpha}o_\beta+o_\alpha\kappa_{2\,\beta})\mathscr{J}^\zeta_{4M\omega_2}(\vec k_1+\vec k_2,z_1,z_2)\nonumber\\&\quad+4MT^{\alpha\dot\alpha}T^{\beta\dot\beta}(z_2\kappa_{1\,\alpha}\iota_\beta-z_1\iota_\alpha\kappa_{2\,\beta})\mathscr{J}^{\bar\zeta}_{4M\omega_2}(\vec k_1+\vec k_2,z_1,z_2)
\nonumber\\&\quad+4MT^{\alpha\dot\alpha}T^{\beta\dot\beta}(-z_2\kappa_{1\,\alpha}o_\beta+\iota_\alpha\kappa_{2\,\beta})\mathscr{J}_{4M\omega_2}^{\zeta\bar\zeta}(\vec k_1+\vec k_2,z_1,z_2)
\nonumber\\&\quad\left.+8M^2T^{\alpha\dot\alpha}\kappa_{1\,\alpha}T^{\beta\dot\beta}\kappa_{2,\beta}\mathscr{K}_{4M\omega_2}(\vec k_1+\vec k_2,z_1,z_2)\right)\,,\label{eq:graviton_integral_decomposition}
\end{align}
\endgroup
with again $n=4M\omega_2$. Thus the computation of all the four amplitudes reduces to the evaluation of \eqref{eq:In-integral}-\eqref{eq:Kn-integral}: to do so, we introduce auxiliary ``master integrals'' that can be evaluated explicitly and used to recover \eqref{eq:In-integral}-\eqref{eq:Kn-integral}.
\subsection{$\mathscr{H}_n$}
Let us start with the computation of \eqref{eq:In-integral}: we first introduce the complex parameters
\be
w=\frac{k^1+\im k^2}{k^3}\,,\qquad \bar w=\frac{k^1-\im k^2}{k^3}
\ee
so that we can parametrize the 3-vectors $\vec x$ and $\vec k$ as
\be
\vec x=\frac{r}{1+|\zeta|^2}(\zeta+\bar\zeta,\im(\bar\zeta-\zeta),1-|\zeta|^2)\,,\qquad\vec k=\frac{1}{2}k^3(w+\bar w,\im(\bar w-w),2)\,.
\ee
Some useful identities are
\be
\vec k\cdot\vec x=\frac{k^3r}{1+|\zeta|^2}(\zeta\bar w+\bar\zeta w+1-|\zeta|^2)\,,\qquad\vec k\,^2=(k^3)^2(1+|w|^2)\,.
\ee
Upon performing the rescaling $r\to r(1+|\zeta|^2)$, the integral $\mathscr{H}_n(\vec k,z_1,z_2)$ reduces to
\be
\mathscr{H}_n(\vec k,z_1,z_2)=2\im\int_0^\infty\d r\,r^2\int_\C\d\zeta\,\d\bar\zeta\,(1+|\zeta|^2)[r(\zeta-z_1)(\bar\zeta z_2+1)]^n\e^{\im k^3r(\zeta\bar w+\bar\zeta w+1-|\zeta|^2)}\,.
\ee
We can now introduce our first master integral as
\be
\hat{\mathscr{H}}(\vec k,z_1,z_2;t)=2\im\int_0^\infty\d r\,r^2\int_\C\d\zeta\,\d\bar\zeta\,(1+|\zeta|^2)\e^{\im k^3r(\zeta\bar w+\bar\zeta w+1-|\zeta|)^2+\im tr(\zeta-z_1)(\bar\zeta z_2+1)}\,.
\ee
$\mathscr{H}_n(\vec k,z_1,z_2)$ can now be recovered as
\be
\mathscr{H}_n(\vec k,z_1,z_2)=\left.(-\im \p_t)^n\hat{\mathscr H}(\vec k,z_1,z_2;t)\right|_{t=0}\,.
\ee
The advantage in computing $\hat{\mathscr{H}}$ instead of evaluating $\mathscr{H}_n$ directly is manifest, as the former is a simple Gaussian integral over the stereographic coordinates: introducing the parameters 
\be
A=tz_2-k^3\,,\quad B=k^3\bar w+t\,,\quad C=k^3w-tz_1z_2\,,\quad D=k^3-tz_1\,,\label{eq:parameters}
\ee
we can express the master integral as
\be
\hat{\mathscr{H}}(\vec k,z_1,z_2;t)=2\im\int_0^\infty\d r\,r^2\int_\C\d\zeta\,\d\bar\zeta\,(1+|\zeta|^2)\e^{\im r(A|\zeta|^2+B\zeta+C\bar\zeta+D)}\,.
\ee
As we will set $t=0$ at the end of the calculation, we are free to think of $t$ to be small enough so that the imaginary part of $tz_2$ does not dominate our $\im\eps$ regulator. 

Performing the Gaussian integral leaves us with a Mellin integral over the radial variable
\be
\hat{\mathscr{H}}(\vec k,z_1,z_2;t)=\frac{4\pi\im}{A^3}\int_0^\infty\d r\,(r(A^2+BC)+\im A)\e^{\im r(AD-BC)/A}\,.
\ee
Since we are interested in taking derivatives of the master integral at $t=0$ and since the function $(AD-BC)/A$ has an imaginary part at least linear in $t$, the implied $\im\epsilon$ prescription allows us to make this residual Mellin integral convergent as well. In this way, we finally find
\be
\begin{aligned}
\hat{\mathscr{H}}(\vec k,z_1,z_2;t)&=-\frac{4\pi\im(A+D)}{(AD-BC)^2}\\&=\frac{4\pi\im tz_{12}}{(\vec k\,^2+tQ)^2}\,,
\end{aligned}
\ee
where in the second line we plugged in the values \eqref{eq:parameters} and defined the quantity
\be
Q=k^3(w-z_1-z_2-\bar wz_1z_2)\,.
\ee
Upon taking $n$ derivatives, we conclude
\be
\mathscr{H}_n(\vec k,z_1,z_2)=4\pi n\,n!\frac{z_{12}(\im Q)^{n-1}}{(\vec k\,^2)^{n+1}}\,.\label{eq:In_evaluated}
\ee
Notice in particular that $\mathscr{H}_n$ has the expected mass dimension: rescaling the radial variable in \eqref{eq:In-integral} to set $|\vec k|$ to 1, we expect $\mathscr{H}_n\sim|\vec k|^{-n-3}$, which is precisely the scaling of \eqref{eq:In_evaluated} once we notice that $Q\sim|\vec k|$.
\subsection{$\mathscr{J}_n$ and $\mathscr{K}_n$}
The computation of \eqref{eq:J0n-integral}-\eqref{eq:Kn-integral} proceeds along the same lines, so we can be brief. After the rescaling $r\to r(1+|\zeta|^2)$, the integrals read
\begin{align}
    \mathscr{J}^0_n(\vec k,z_1,z_2)&=2\im\int_0^\infty\d r\,r^2\int_\C\d\zeta\,\d\bar\zeta\,[r(\zeta-z_1)(\bar \zeta z_2+1)]^{n-1}\e^{\im\vec k\cdot\vec x}\,,\\
    \mathscr{J}^\zeta_n(\vec k,z_1,z_2)&=2\im\int_0^\infty\d r\,r^2\int_\C\d\zeta\,\d\bar\zeta\,\zeta\,[r(\zeta-z_1)(\bar \zeta z_2+1)]^{n-1}\e^{\im\vec k\cdot\vec x}\,,\\
    \mathscr{J}^{\bar\zeta}_n(\vec k,z_1,z_2)&=2\im\int_0^\infty\d r\,r^2\int_\C\d\zeta\,\d\bar\zeta\,\bar\zeta\,[r(\zeta-z_1)(\bar \zeta z_2+1)]^{n-1}\e^{\im\vec k\cdot\vec x}\,,\\
    \mathscr{J}^{\zeta\bar\zeta}_n(\vec k,z_1,z_2)&=2\im\int_0^\infty\d r\,r^2\int_\C\d\zeta\,\d\bar\zeta\,|\zeta|^2\,[r(\zeta-z_1)(\bar \zeta z_2+1)]^{n-1}\e^{\im\vec k\cdot\vec x}\,,\\
    \mathscr{K}_n(\vec k,z_1,z_2)&=2\im\int_0^\infty\d r\,r\int_\C\d\zeta\,\d\bar\zeta\,[r(\zeta-z_1)(\bar \zeta z_2+1)]^{n-1}\e^{\im\vec k\cdot\vec x}\,,
\end{align}
The relevant master integrals are thus
\begin{align}
    \hat{\mathscr{J}}^0(\vec k,z_1,z_2;t)&=2\im\int_0^\infty\d r\,r^2\int_\C\d\zeta\,\d\bar\zeta\,\e^{\im r(A|\zeta|^2+B\zeta+C\bar\zeta+D)}\,,\\
    \hat{\mathscr{J}}^\zeta(\vec k,z_1,z_2;t)&=2\im\int_0^\infty\d r\,r^2\int_\C\d\zeta\,\d\bar\zeta\,\zeta\e^{\im r(A|\zeta|^2+B\zeta+C\bar\zeta+D)}\,,\\
    \hat{\mathscr{J}}^{\bar\zeta}(\vec k,z_1,z_2;t)&=2\im\int_0^\infty\d r\,r^2\int_\C\d\zeta\,\d\bar\zeta\,\bar\zeta\e^{\im r(A|\zeta|^2+B\zeta+C\bar\zeta+D)}\,,\\
    \hat{\mathscr{J}}^{\zeta\bar\zeta}(\vec k,z_1,z_2;t)&=\,,2\im\int_0^\infty\d r\,r^2\int_\C\d\zeta\,\d\bar\zeta\,|\zeta|^2\e^{\im r(A|\zeta|^2+B\zeta+C\bar\zeta+D)}\\
    \hat{\mathscr{K}}(\vec k,z_1,z_2;t)&=2\im\int_0^\infty\d r\,r\int_\C\d\zeta\,\d\bar\zeta\,\e^{\im r(A|\zeta|^2+B\zeta+C\bar\zeta+D)}\,,
\end{align}
and now the desired integrals can be recovered by taking $n-1$ derivatives only, e.g.
\be
\mathscr{J}^0_n(\vec k,z_1,z_2)=\left.(-\im \p_t)^{n-1}\hat{\mathscr{J}}^0(\vec k,z_1,z_2;t)\right|_{t=0}\,.
\ee
After performing the Gaussian integrals over $\zeta$, $\bar\zeta$ and the Mellin integral over $r$, we find
\begin{align}
    \hat{\mathscr{J}}^0(\vec k,z_1,z_2;t)&=\frac{4\pi\im(k^3-tz_2)}{(\vec k\,^2+tQ)^2}\,,\\
    \hat{\mathscr{J}}^\zeta(\vec k,z_1,z_2;t)&=\frac{4\pi\im(k^3w-tz_1z_2)}{(\vec k\,^2+tQ)^2}\,,\\
    \hat{\mathscr{J}}^{\bar\zeta}(\vec k,z_1,z_2;t)&=\frac{4\pi\im(k^3\bar w+t)}{(\vec k\,^2+tQ)^2}\,,\\
    \hat{\mathscr{J}}^{\zeta\bar\zeta}(\vec k,z_1,z_2;t)&=-\frac{4\pi\im(k^3-tz_1)}{(\vec k\,^2+tQ)^2}\,,\\
    \hat{\mathscr{K}}(\vec k,z_1,z_2;t)&=\frac{4\pi}{\vec k\,^2+tQ}\,,
\end{align}
so that
\begin{align}
    \mathscr{J}^0_n(\vec k,z_1,z_2)&=-4\pi(n-1)!\frac{(\im Q)^{n-2}}{(\vec k\,^2)^n}\left(\frac{nk^3Q}{\vec k\,^2} +(n-1)z_2\right)\,,\\
    \mathscr{J}^\zeta_n(\vec k,z_1,z_2)&=-4\pi(n-1)!\frac{(\im Q)^{n-2}}{(\vec k\,^2)^n}\left(\frac{nk^3wQ}{\vec k\,^2}+ (n-1)z_1z_2\right)\,,\\
    \mathscr{J}^{\bar\zeta}_n(\vec k,z_1,z_2)&=-4\pi(n-1)!\frac{(\im Q)^{n-2}}{(\vec k\,^2)^n}\left(\frac{nk^3\bar wQ}{\vec k\,^2} -(n-1)\right)\,,\\
    \mathscr{J}^{\zeta\bar\zeta}_n(\vec k,z_1,z_2)&=4\pi(n-1)!\frac{(\im Q)^{n-2}}{(\vec k\,^2)^n}\left(\frac{nk^3Q}{\vec k\,^2}+ (n-1)z_1\right)\,,\\
    \mathscr{K}_n(\vec k,z_1,z_2)&=4\pi(n-1)!\frac{(\im Q)^{n-1}}{(\vec k\,^2)^n}\,.
\end{align}

To get the desired expressions for the amplitudes, we can simplify the formulae for the integrals further using on-shell kinematics, i.e. using the fact that $\vec k=\vec k_1+\vec k_2$ and expressing $Q$, $k^3$, $\vec k\,^2$, $w$, and $\bar w$ in terms of $z_1$, $z_2$, $\omega_1$, and $\omega_2$. On the support of $\omega_1+\omega_2=0$, the relevant expressions are
\begin{align}
    \vec k\,^2&=-\frac{4z_{12}\tilde z_{12}\omega_2^2}{(1+z_1\tilde z_1)(1+z_2\tilde z_2)}\,,\\
    k^3&=-2\im\omega_2\frac{z_1\tilde z_1-z_2\tilde z_2}{(1+z_1\tilde z_1)(1+z_2\tilde z_2)}\,,\\k^3w&=-2\im\omega_2\frac{-z_{12}+\tilde z_{12}z_1z_2}{(1+z_1\tilde z_1)(1+z_2\tilde z_2)}\,,\\k^3\bar w&=-2\im\omega_2\frac{-\tilde z_{12}+z_{12}\tilde z_1\tilde z_2}{(1+z_1\tilde z_1)(1+z_2\tilde z_2)}\,,\\Q&=-2\im\omega_1z_{12}\,.
\end{align}
With these, we immediately obtain the vanishing of the scalar and photon amplitudes: for example, in the photon case the amplitude is
\begin{multline}
\mathcal{A}(1^{-1}_{\omega=-},2^{-1}_{\omega=+})=8\pi^2\delta_{-\omega_1,\omega_2}\langle 1\,2\rangle^2(4M\omega_2)!\frac{z_{12}^{4M\omega_2}}{|\vec k_1+\vec k_2|^{8M\omega_2+2}}\\\times\left(4M\omega_2(2\omega_1)^{4M\omega_2-1}+2M(2\omega_1)^{4M\omega_2}\right)\,,
\end{multline}
and the bracket vanishes on the support of $\omega_1+\omega_2=0$. An analogous cancellation, although slightly more involved, occurs for $\mathcal{A}(1^0_{e=-},2^0_{e=+})$. For the 2-point gluon amplitude, we find
\be
\mathcal{A}(1^{-1}_{e=-},2^{-1}_{e=+})=-8\pi^2 \delta(\omega_1+\omega_2)\delta_{-e_1,e_2}\langle 1\,2\rangle^2e_2\,(e_2)!\frac{(2\omega_1)^{2e_2-1}z_{12}^{2e_2}}{|\vec k_1+\vec k_2|^{4e_2+2}}\,,
\ee
and upon expressing $z_{12}=\langle 1\,2\rangle/(\kappa_1^0\kappa_2^0)$ in terms of the spinor-helicity variables, we recover \eqref{eq:integrated-2-point-gluon}. The computation for the 2-point graviton amplitude proceeds along the same lines, except that the technical details are more involved. Upon inserting \eqref{eq:In-integral}-\eqref{eq:Kn-integral} into \eqref{eq:graviton_integral_decomposition}, we find
\begin{multline}
\mathcal{A}(1^{-2}_{\omega=-},2^{-2}_{\omega=+})=-8\pi^2\frac{\langle 1\,2\rangle^3\eta_{1\,\dot\alpha}\eta_{2\,\dot\beta}}{[\eta_1\,1][\eta_2\,2]}\delta_{-\omega_1,\omega_2} (4M\omega_2-1)!\frac{(2\omega_1)^{4M\omega_2-1}z_{12}^{4M\omega_2+2}}{|\vec k_1+\vec k_2|^{8M\omega_2+2}}M^2\times\\\times\frac{-2\omega_2^2}{(1+z_1\tilde z_1)(1+z_2\tilde z_2)}\left(\begin{array}{cc}
     1&\tilde z_2\\\tilde z_1&\tilde z_1\tilde z_2
\end{array}\right)^{\dot\alpha\dot\beta}\,.
\end{multline}
We can recognize the tensor structure in the second line as the tensor product $\tilde\kappa_1^{\dot\alpha}\tilde\kappa_2^{\dot\beta}$, as we expect it to be because the amplitude ought to be gauge invariant, i.e., independent of the choice of the reference spinors $\eta_{1\,\dot\alpha}$ and $\eta_{2\,\dot\alpha}$. With this simplification and expressing again $z_{12}/(\kappa_1^0\kappa_2^0)$ as $\langle 1\,2\rangle$, we finally recover \eqref{eq:integrated-2-point-graviton}.

\bibliographystyle{JHEP}
\bibliography{sdc}

\end{document}